\documentclass{PoS}
\usepackage{epsfig}

\title{Forward-backward and $CP$-violating asymmetries in rare semileptonic 
       and radiative leptonic $B$-decays}

\ShortTitle{Asymmetries in rare $B$-decays}

\author{\speaker{Nikolai Nikitin}\\
        D.~V.~Skobeltsyn Institute of Nuclear Physics, Moscow State University\\
        E-mail: \email{nnikit@mail.cern.ch}}

\author{Irina Balakireva \\
        D.~V.~Skobeltsyn Institute of Nuclear Physics, Moscow State University\\
        E-mail: \email{iraxff@mail.ru}}

\author{Dmitri Melikhov \\
        D.~V.~Skobeltsyn Institute of Nuclear Physics, Moscow State University\\
        E-mail: \email{dmitri\_melikhov@gmx.de}}

\abstract{We study the forward-backward and the $CP$-violating asymmetries (both 
time-independent and time-dependent) in rare semileptonic and radiative leptonic 
$B_{d,s}$-decays and investigate the sensitivity of these asymmetries to the extensions 
of the Standard model.}

\FullConference{The XIXth International Workshop on High Energy Physics and Quantum Field Theory, QFTHEP2010\\
		September 08-15, 2010\\
		Golitsyno, Moscow, Russia}

\begin{document}

\section*{Introduction}

Rare $B$-decays induced by flavor-changing neutral currents provide a valuable possibility 
of an indirect search of physics beyond the standard model (SM). $CP$-violation in beauty 
sector has been measured for the first time at $B$-factories BaBar and Belle in two-hadron 
$B$-meson decays \cite{CP-in-B}. Other interesting reactions, where $CP$-violating effects 
may be studied, are rare semileptonic and radiative leptonic decays induced by 
$b \to (d, s)\,(\gamma,\, \ell^+ \ell^-)$ transitions (see, e.g., \cite{stone} and refs 
therein). 

Obviously, an experimental study of $CP$-violating observables requires greater samples 
of beauty hadrons than those provided by the B-factories: since $CP$-violation effects 
in beauty sector are of order $10^{-3}$, one needs $B$-meson data samples exceeding those 
from $B$-factories by at least two orders of magnitude. One expects such data samples of 
beauty particles at the LHC: e.g., the detector LHCb is expected to register about 
$10^{12}$ $b\bar b$-pairs per year, approximately by four orders of magnitude more than 
a yearly yield of $b\bar b$-pairs at the $B$-factories \cite{LHCb}. 

The time-independent $CP$-asymmetries in rare semileptonic decays were considered first 
for inclusive $B$-decays, i.e., for the process $b \to d \ell^+ \ell^-$ \cite{krugerCP}. 
An asymmetry of the order of a few percent in $b \to d \ell^+ \ell^-$ has been predicted; 
for the $b \to s \ell^+ \ell^-$-transitions the asymmetry is expected to be much smaller. 
Following \cite{krugerCP}, the time-independent $CP$-asymmetries in exclusive $B$-decays 
for several extentions of the SM have been analyzed \cite{CPextra,hiller}.

The time-dependent CP-asymmetries \cite{dunietz} were studied for the case of rare 
semileptonic $B_{s}\to \phi \ell^+ \ell^-$--decays in Ref.~\cite{hiller}. 

This talk presents the results of our analysis \cite{bmnt}
of the asymmetries (forward-backward, time-in\-de\-pen\-dent and 
time-dependent $CP$-violating asymmetries) in rare semileptonic and radiative leptonic 
$B_{s,d}\to (V,\gamma)  \ell^+ \ell^-$-decays. 
We make use of the technique of helicity amplitudes developed in \cite{hagiwara,soni}. 


\section{Theoretical overview}

The effective Hamiltonian describing the $b\to s \ell^+\ell^-$ transition in the SM has the 
form \cite{wc}\footnote{We use the following conventions: 
$\gamma^5=i\gamma^0\gamma^1\gamma^2\gamma^3$, 
$\sigma_{\mu\nu}=\frac{i}{2}[\gamma_{\mu},\gamma_{\nu}]$, 
$\varepsilon^{0123}=-1$, $e=\sqrt{4\pi\alpha_{\rm em}} > 0$.} 
\begin{eqnarray}
\label{SMb2qll}
&&H_{\rm eff}^{\rm SM}
(b\to s\ell^+\ell^-) = \,\frac{G_{F}}{\sqrt2}\, \frac{\alpha_{em}}{2\pi}\,
V_{tb}V^*_{ts}\,
\left[\,-2\, {\frac{C_{7\gamma}(\mu)}{q^2}} \cdot 
\bar s\, i\sigma_{\alpha\beta}q^\beta\left\{m_b(1+\gamma^5)+ m_s(1-\gamma^5) \right\} b
\cdot 
\bar \ell\gamma^{\alpha}\ell\right. \nonumber \\
&&\qquad\qquad\left. +\, 
C_{9V}^{\rm eff\,(s)}(\mu,\, q^2)\cdot \bar s\gamma_{\alpha}\left (1-\gamma^5\right ) b \cdot \,\bar \ell\gamma^{\alpha}\ell \, 
+\,
C_{10A}(\mu)\cdot \bar s\gamma_{\alpha}\left (1-\gamma^5\right ) b \cdot \,{\bar \ell}\gamma^{\alpha}\gamma^5\ell \right], 
\end{eqnarray} 
where $m_b$ ($m_s$) is the $b$ ($s$)--quark mass, $V_{ij}$ are the elements of the CKM matrix,  
$\mu$ is the re\-nor\-ma\-li\-za\-ti\-on scale, and $q$ is the momentum of the 
$\ell^+\ell^-$ pair. The corresponding expression for the case of the $b\to d$ transition 
is self-evident.

The Wilson coefficient $C_{9V}^{\rm eff}(\mu,\, q^2)$ contains the contributions of the 
virtual $\bar uu$ and $\bar cc$ pairs, which involve the integration over short and long 
distances. The long-distance effects are described by the neutral vector-meson resonances 
$\rho$, $\omega$, and $J/\psi$, $\psi'$. We make use of the parameterization of 
$C_{9V}^{\rm eff}(\mu,\, q^2)$ from \cite{krugerCP} where the resonance contributions are 
modeled in a gauge-invariant way \cite{melikhovplb}.

The effective Hamiltonian for the $\bar b\to \bar s\ell^+\ell^-$ transition is obtained 
from (\ref{SMb2qll}) by interchanging $b$ and $s$ fields, i.e., by replacing 
$V_{qb}V^*_{qs}\to V^*_{qb}V_{qs}$, $\bar s \to \bar b$, $b \to s$, $m_b\leftrightarrow m_s$, 
$q \to -q$.

The form factors for rare semileptonic transitions 
$\bar B(p_1,\, M_1)\to \bar V(p_2,\, M_2,\,\varepsilon)$ are defined in the standard way 
\cite{BDandK}, where $q = p_1 - p_2$. The $B\to\gamma$ amplitudes are parametrized as 
follows \cite{mk}, where $q = p - k$, $k^2=0$, $p^2=M_1^2$.


\section{Forward-backward and CP-violating asymmetries}

The lepton forward-backward asymmetry $A_{FB}(\hat s)$ is one of the differential 
distributions re\-la\-ti\-ve\-ly stable with respect to QCD uncertainties and sensitive to 
the new physics effects \cite{mns2}. Therefore it has been extensively studied both 
theoretically and experimentally \cite{BDandK}. We make use of the following definition 
of $A_{FB}(\hat s)$ for $\bar B\to \bar f$ decays 
\begin{eqnarray}
\label{Afb}
A_{FB}(\hat s)\, =\,
\frac{\displaystyle\int_0^1 d\cos\theta\,\frac{d^2\Gamma (\bar B \to \bar f)}
               {d \hat s\, d\cos\theta}\, -\,
      \int_{-1}^0 d\cos\theta\,\frac{d^2\Gamma (\bar B \to \bar f)}
               {d \hat s\, d\cos\theta}
     }
     {\displaystyle\frac{d\Gamma (\bar B \to \bar f)}{d\, \hat s}
     },
\end{eqnarray}
where $f=V l^+ l^-$ for rare semileptonic decay and $f=\gamma l^+ l^-$ for rare radiative 
decay; $\hat s\equiv s/M_B^2$, $\sqrt{s}$ being the dilepton invariant mass. 
The asymmetry is calculated in the rest frame of the lepton pair, and the angle
$\theta$ is defined as the angle between the $\bar B$-meson 3-momentum and the 3-momentum 
of the outgoing negative-charged lepton, $l^-$.

Equivalently, for $B\to f$ one defines 
\begin{eqnarray}
\label{Afb1}
A_{FB}(\hat s)\, =\,
\frac{\displaystyle\int_0^1 d\cos\theta_+\,\frac{d^2\Gamma (B \to f)}
               {d \hat s\, d\cos\theta_+}\, -\,
      \int_{-1}^0 d\cos\theta_+\,\frac{d^2\Gamma (B \to f)}
               {d \hat s\, d\cos\theta_+}
     }
     {\displaystyle\frac{d\Gamma (B \to f)}{d\, \hat s}
     }
\end{eqnarray}
where $\theta_+$ is the angle between the $B$-meson 3-momentum and the 3-momentum of the 
outgoing positive-charged lepton, $l^+$. If CP-violating effects are neglected, both 
asymmetries (\ref{Afb}) and (\ref{Afb1}) are equal to each other.

Time-dependent CP-violating asymmetry is defined in the $B$-meson rest frame as follows 
\cite{waldi}
\begin{equation}
\label{ACPtau-def}
A_{CP}^{B_q\to f}(\tau)\, =\,
\frac{\displaystyle\frac{d\Gamma (\bar B^0_q\to f)}{d\tau} -\displaystyle\frac{d\Gamma (B^0_q\to f)}{d\tau}}
     {\displaystyle\frac{d\Gamma (\bar B^0_q\to f)}{d\tau}+\displaystyle\frac{d\Gamma (B^0_q\to f)}{d\tau}}, 
\end{equation}
where $f$ is the common final state for $B^0$ and $\bar B^0$ decays. In this case a 
pronounced CP violation is expected in interference between the oscillation and decay 
amplitudes \cite{sanda}. 
Time-independent $CP$-asymmetry may be represented as follows: 
\begin{eqnarray}
\label{ACPs-equation}
A^{B_q \to f}_{CP}(\hat s)\, =\, 
\frac{\displaystyle{\frac{d\Gamma(\bar B_q\,\to\, f)}{d\hat s}} - \frac{d\Gamma(B_q\,\to\, f)}{d\hat s}}
     {\displaystyle{\frac{d\Gamma(\bar B_q\,\to\, f)}{d\hat s}} + \frac{d\Gamma(B_q\,\to\, f)}{d\hat s}}. 
\end{eqnarray}


\section{Numerical results}

We are going to apply now the formulas derived above and to provide numerical results for 
the asymmetries. We use the following numerical parameters: 

(i) Table \ref{table:Bparam} summarizes the parameters of the $B^0_{d, s}$-oscillations 
which we use for our numerical estimates. 
\begin{table}[bt]
\begin{center}
\begin{tabular}{|l|c|c|}
\hline 
$B$-meson parameters                 & $B^0_d$        & $B^0_s$  \\
\hline 
B-meson mass $M_1$ (GeV)                   & $5.28$         & $5.37$   \\
\hline 
Width $\Gamma$ (ps$^{-1}$)                 & $0.65$         & $0.67  $ \\
\hline 
Mass difference $\Delta m$ (ps$^{-1}$)       & $0.507$        & $17.77$   \\
\hline 
Width difference $\Delta\Gamma$ (ps$^{-1}$)  & $0.005  $      & $0.1$    \\
\hline
\end{tabular}
\caption{Parameters of $B^0_{d,\, s}$-oscillations  \cite{stone,pdg2008,mxz2008}
\label{table:Bparam}}
\end{center}
\end{table}

(ii) The Wilson coefficients for the SM are evaluated at $\mu=5$ GeV \cite{wc}
for $C_2(M_W) = -1$:  $C_1(\mu)=\, 0.241$, $C_2(\mu)=\, -1.1$,
$a_1(\mu) =\, -0.126$, $C_{7\gamma}(\mu)=\, 0.312$, $C_{9V}(\mu)=\, -4.21$ 
and $C_{10A}(\mu)=4.64$. Respectively, we use the running quark masses in the 
$\overline{\rm MS}$ scheme at the same scale: $m_b=4.2$ GeV, $m_s=60-80$ MeV, and 
the $d$-quark mass is neglected. For the coefficient $C_{eV}^{\rm eff}$ we use the 
model proposed in \cite{krugerCP} which takes into account resonances in a 
gauge-invariant way. 

For the CKM matirx elements we use the values reported in the 2008 edition of PDG 
\cite{pdg2008}: $A=0.814$, $\lambda=0.226$, $\bar \rho=0.135$, $\bar \eta=0.35$. 

(iii) We make use of the form factor parameterizations for rare semileptonic decays 
from \cite{ms} and for rare radiative decays from \cite{mk}. The accuracy of 
these predictions for the form factors is expected to be at the level of 10-15\%, 
which influences strongly the predictions for the decay rates. However, the form 
factor uncertainties cancel to a large extent in the asymmetries which therefore can be 
predicted with a few percent accuracy \cite{mns,mns2}. For the decay constants of the 
$B$-mesons we use the values $f_B= 220\pm 20$ MeV and $f_{B_s}= 240\pm 20$ MeV. 

(iv) A cut on the Bremsstrahlung photon spectrum at 20 MeV in the $B$-meson rest frame 
is applied. This corresponds to the expected level of the photon energy resolution of 
the LHCb detector. 

\subsection{Forward-backward asymmetry}
\begin{figure}[tb]
\begin{center}
\begin{tabular}{cc}

\includegraphics[width=5.8cm]{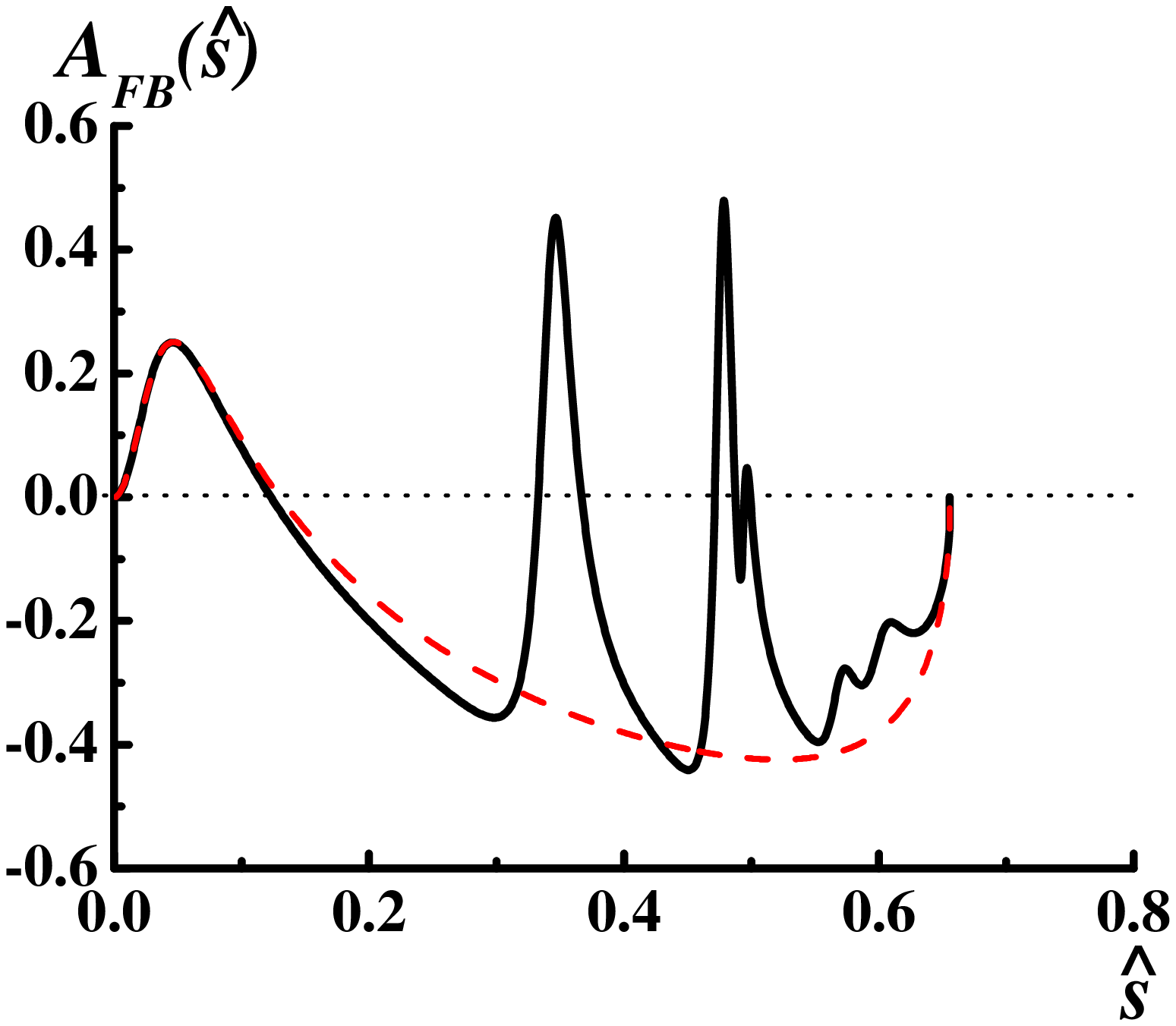} 
& 
\includegraphics[width=5.8cm]{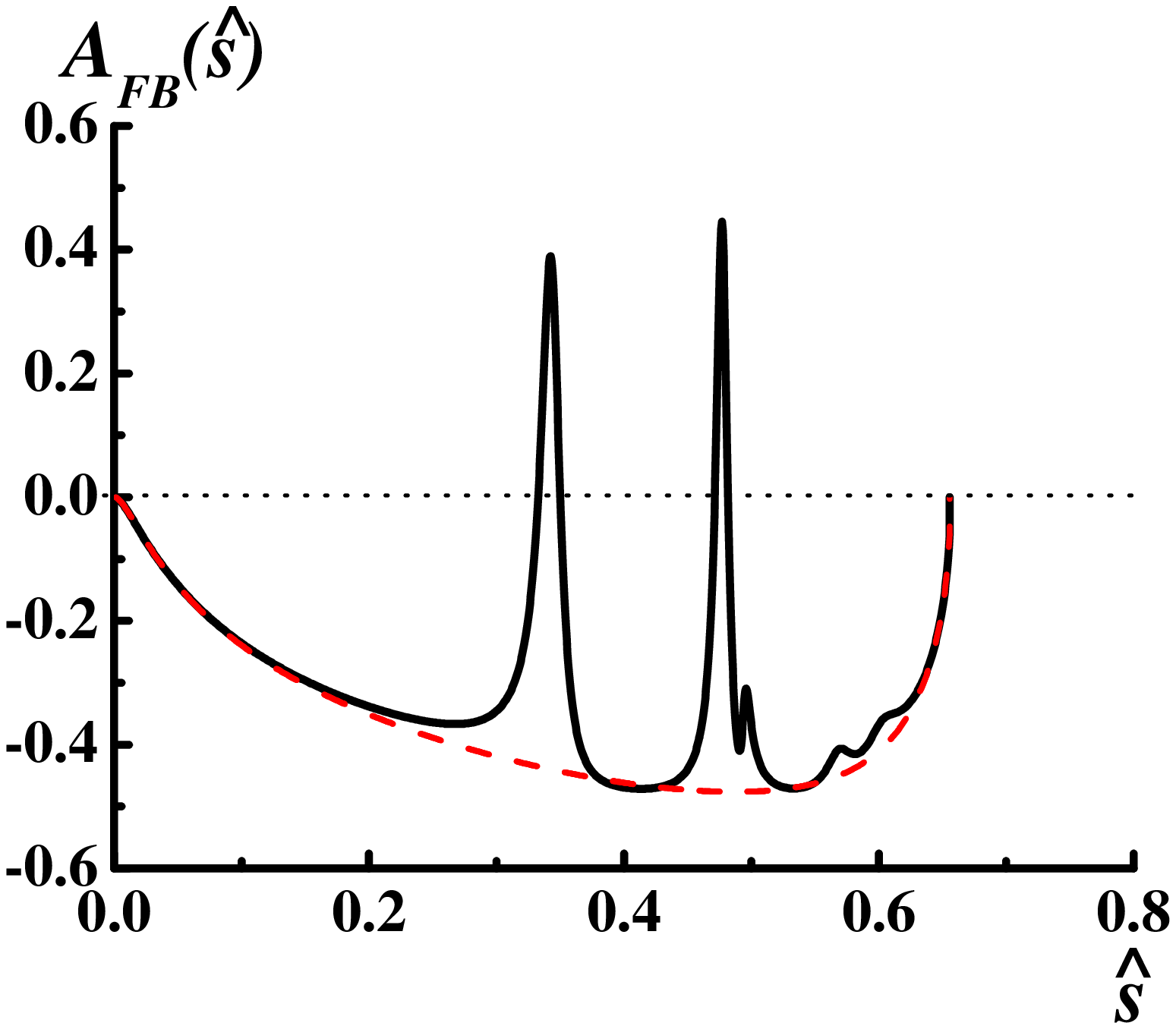} 
\\
(a) & (b)
\\
\includegraphics[width=5.8cm]{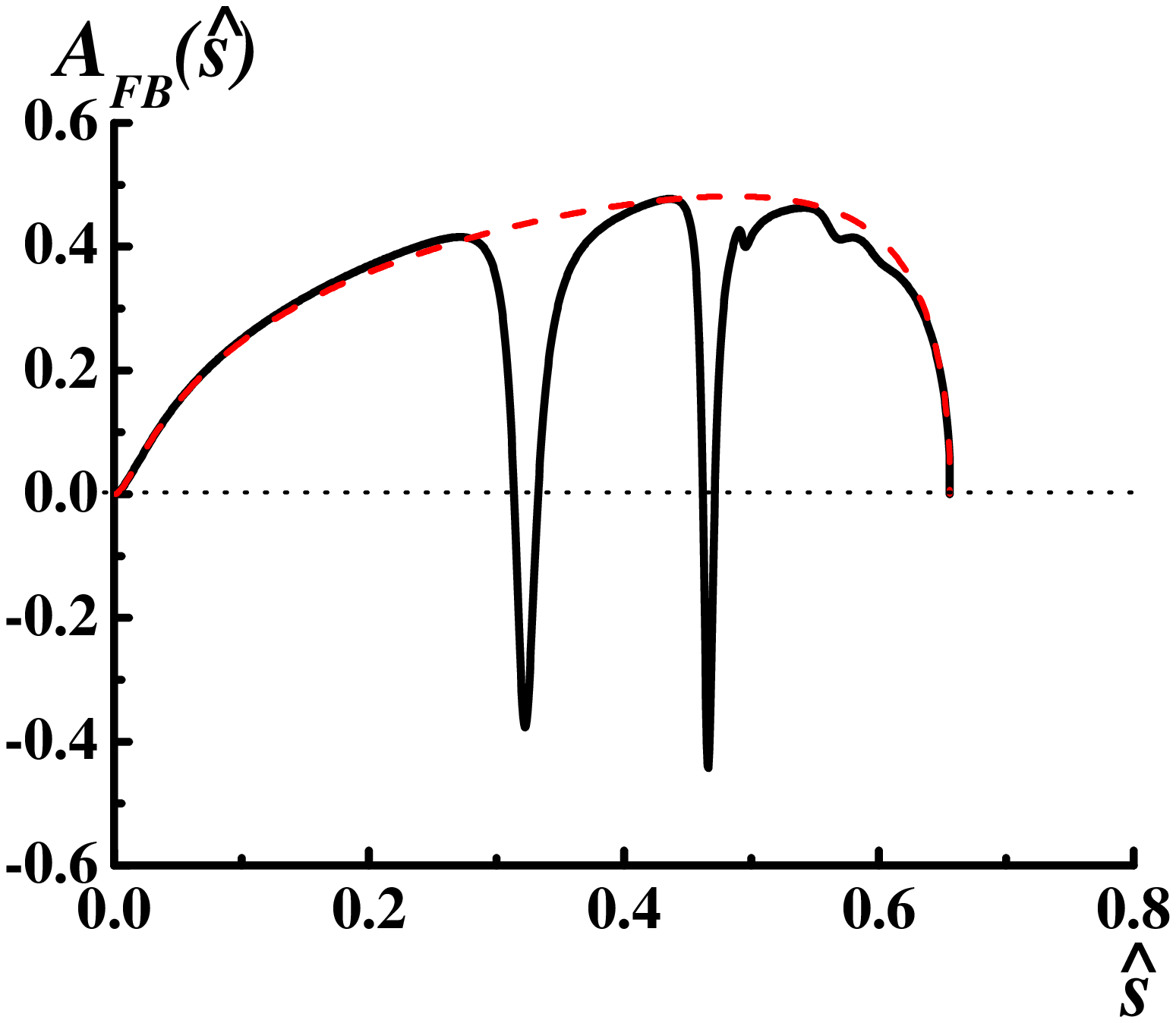} 
& 
\includegraphics[width=5.8cm]{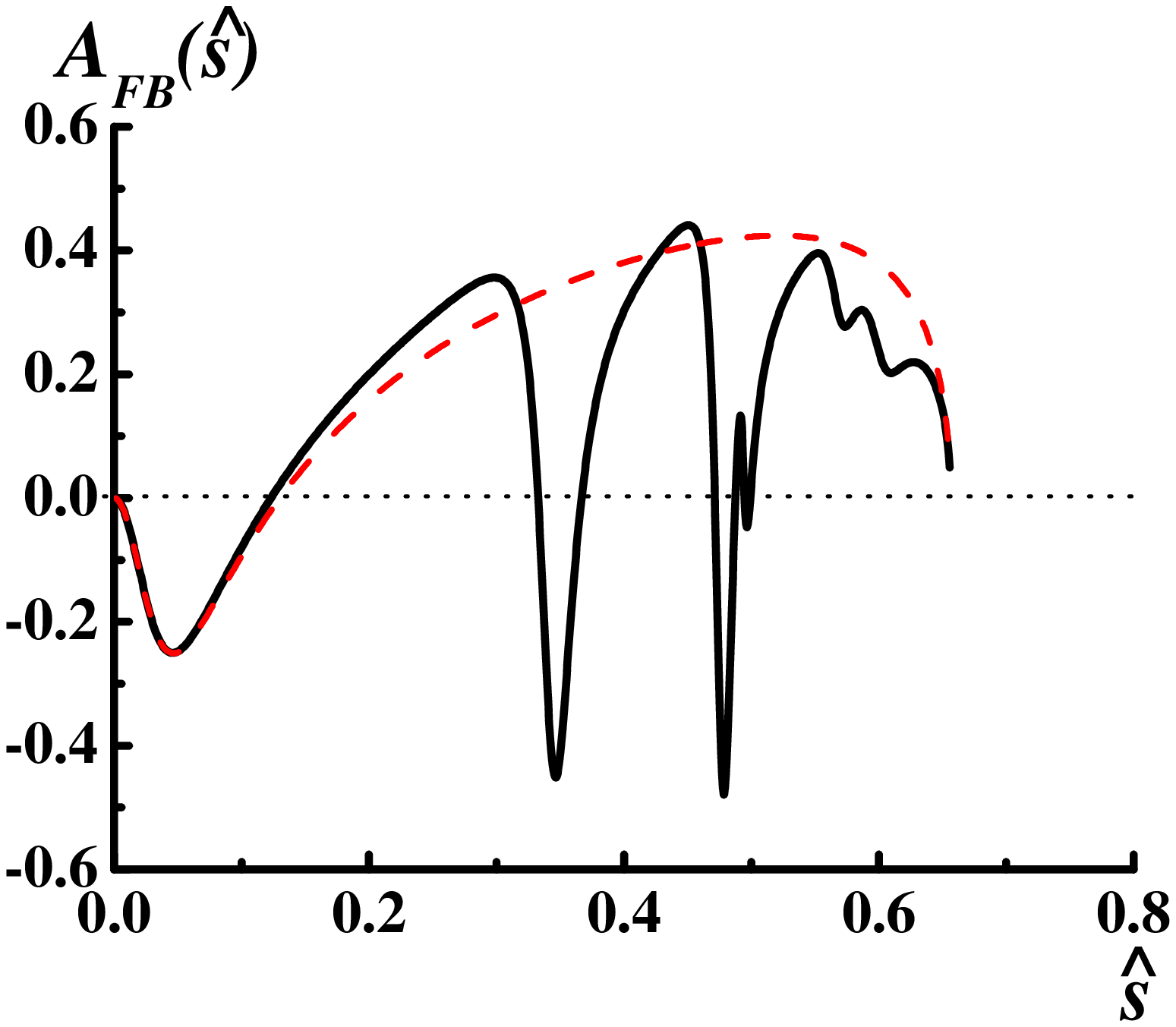}
\\
(c) & (d)
\end{tabular}
\end{center}
\caption{\label{Fig:1}
$A_{FB}$ for rare semileptonic $\bar B_s\to \phi\mu^+\mu^-$ decays: (a) in the SM;  
(b) For $C_{7\gamma}=-C^{\rm SM}_{7\gamma}$, (c) For $C_{9V}=-C^{\rm SM}_{9V}$, (d) For $C_{10A}=-C^{\rm SM}_{10A}$. 
Solid line (black): the full asymmetry which takes into account the $J/\psi$, $\psi'$, etc contributions. 
Dashed line (red): the non-resonant asymmetry.}
\end{figure}

\begin{figure}[bt]
\begin{center}
\begin{tabular}{cc}
\includegraphics[width=5.8cm]{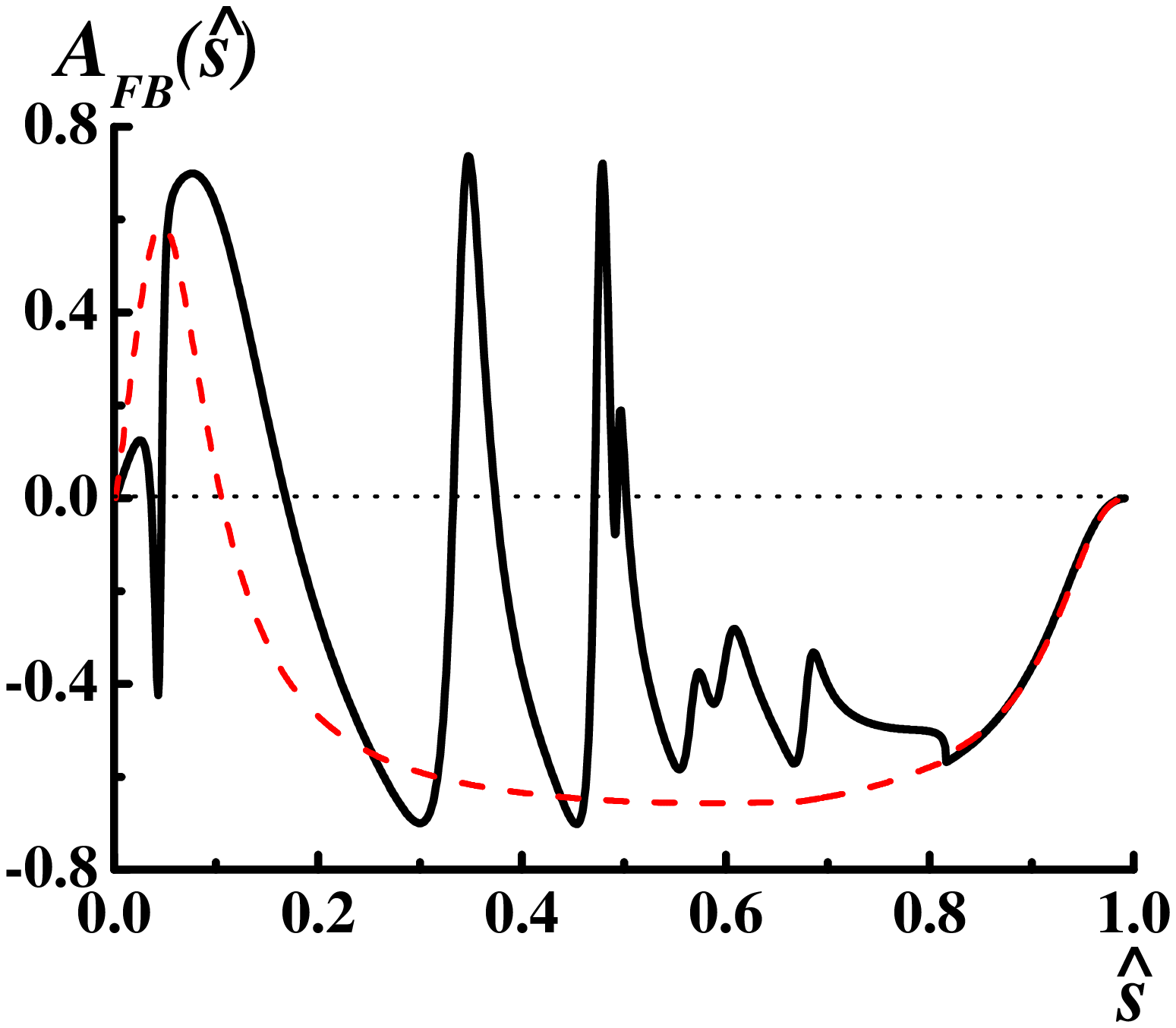} 
& 
\includegraphics[width=5.8cm]{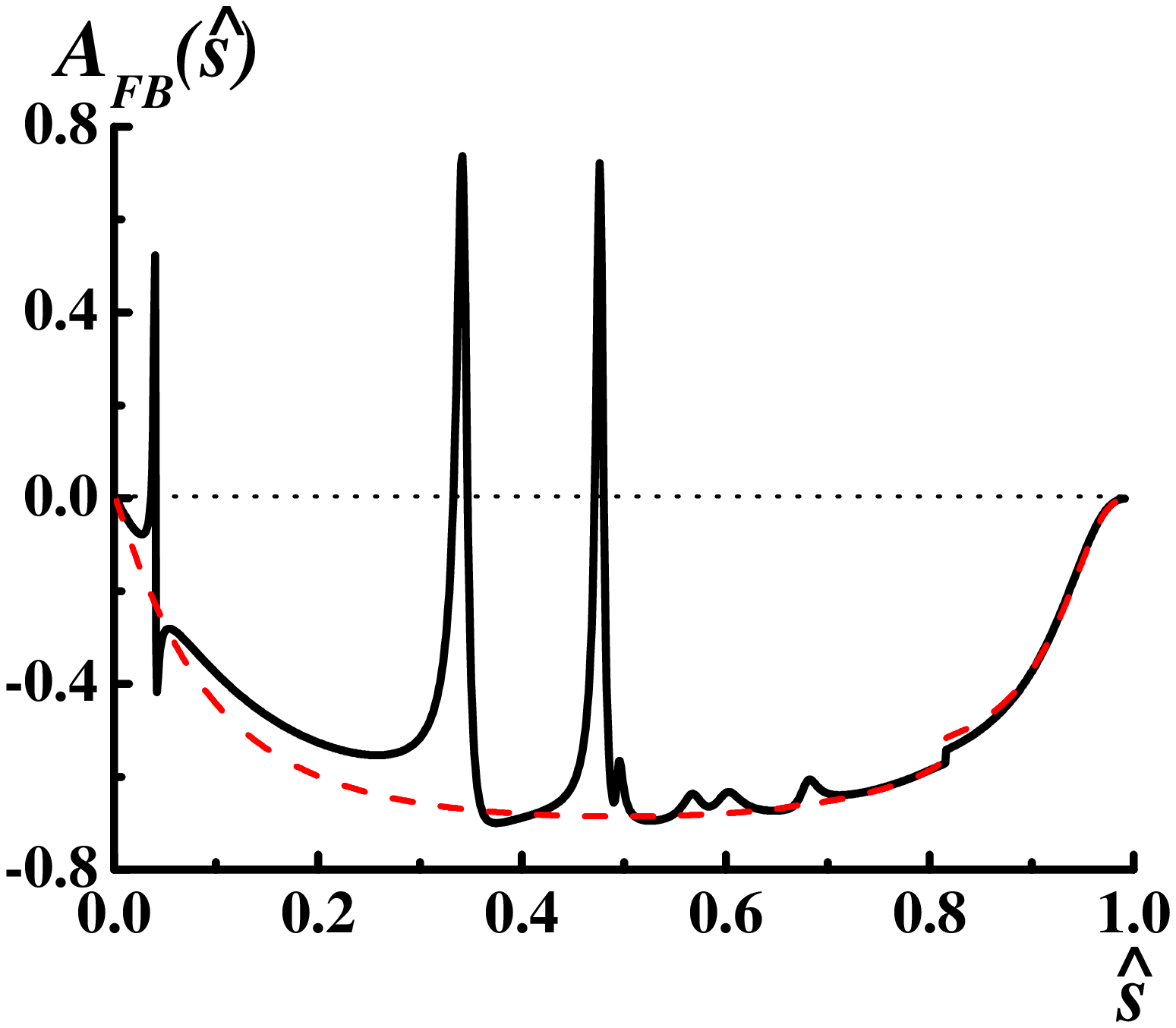} 
\\
(a) & (b)
\\
\includegraphics[width=5.8cm]{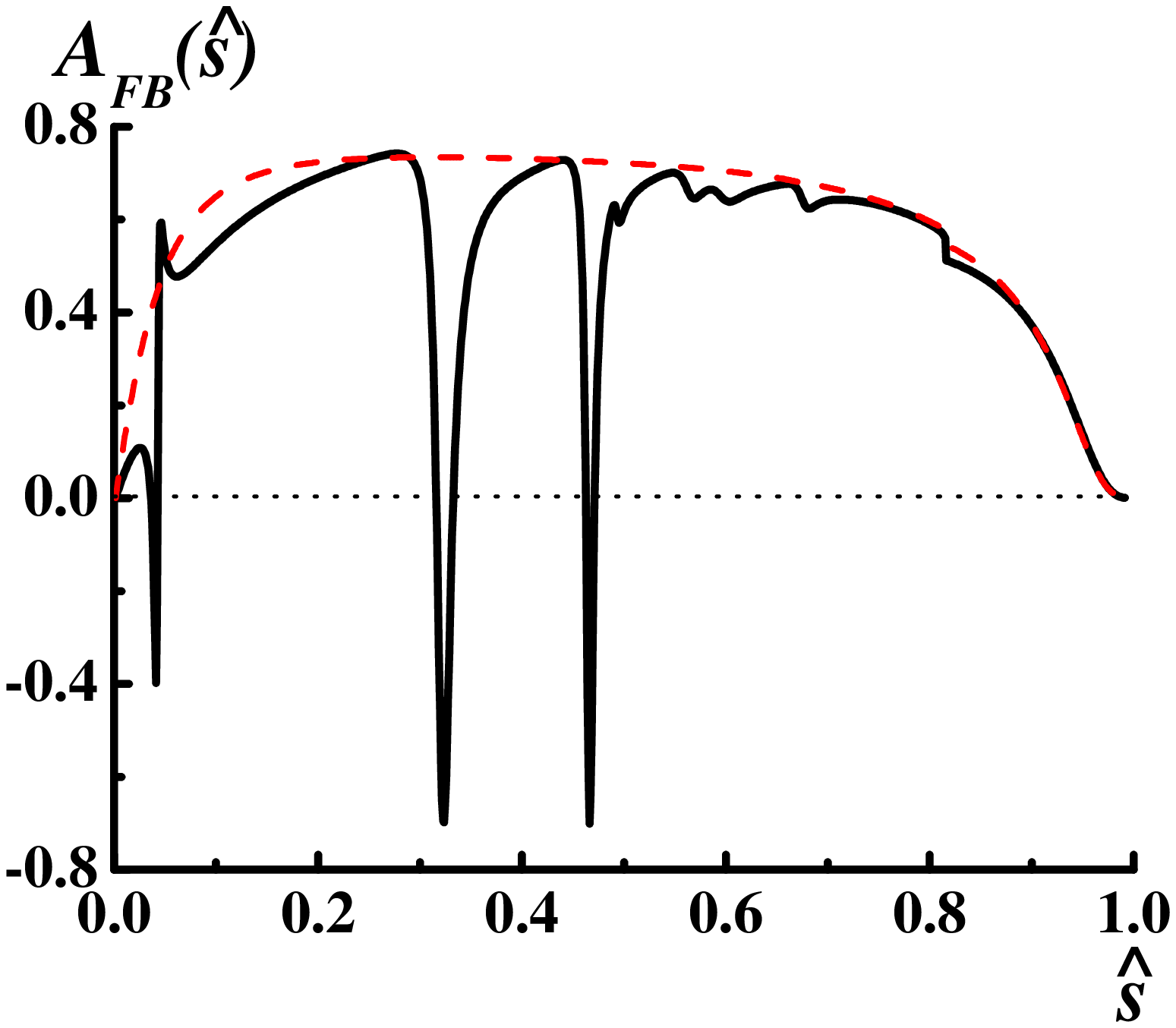} 
& 
\includegraphics[width=5.8cm]{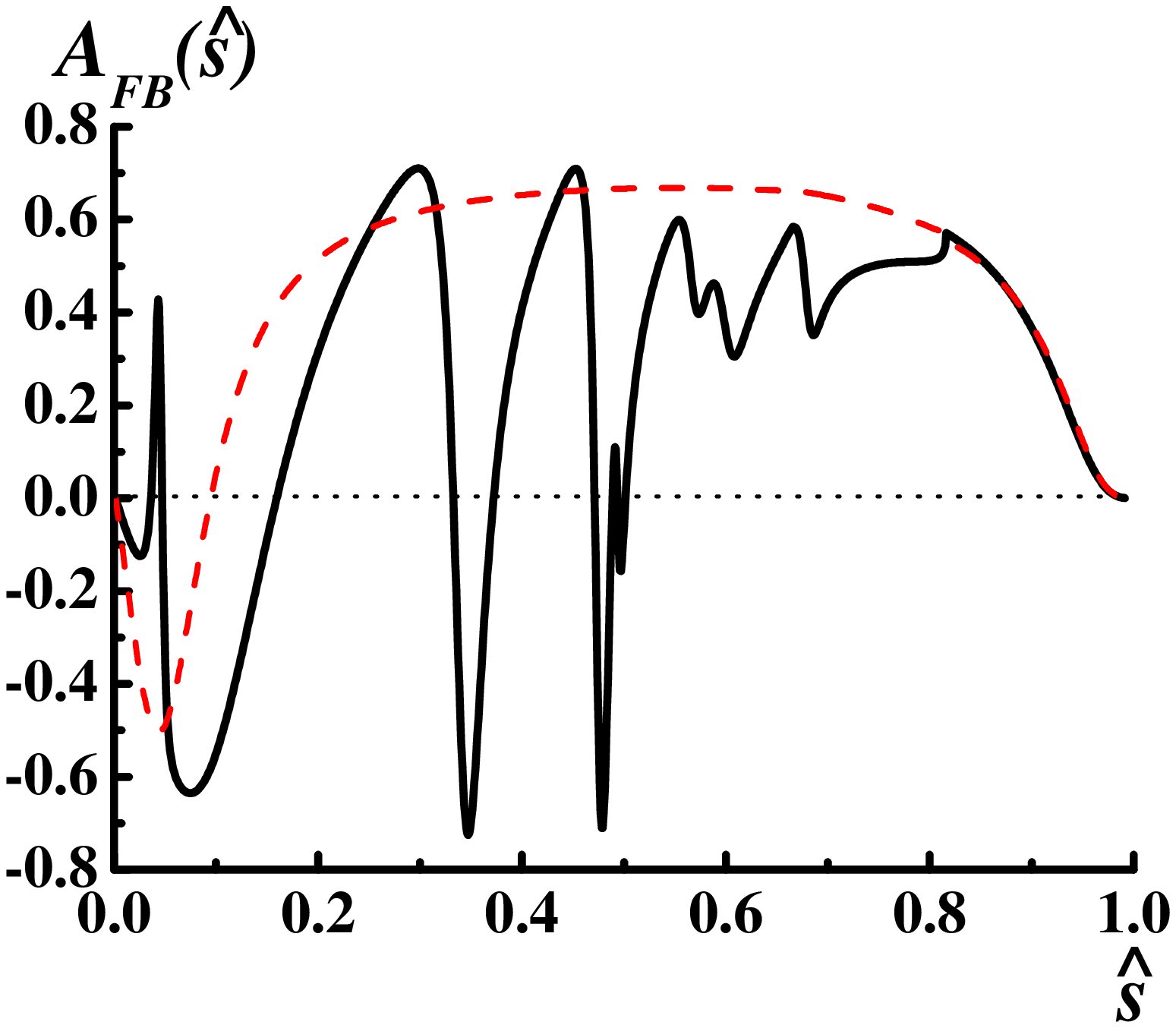}
\\
(c) & (d)
\end{tabular}
\end{center}
\caption{\label{Fig:2}
$A_{FB}(\bar s)$ for $\bar B_s\to \gamma\mu^+\mu^-$ decays: 
(a) In the SM.   
(b) For $C_{7\gamma}=-C^{\rm SM}_{7\gamma}$. 
(c) For $C_{9V}=-C^{\rm SM}_{9V}$. 
(d) For $C_{10A}=-C^{\rm SM}_{10A}$. 
Solid line (black): the asymmetry calculated for the full amplitude of Ref.~\cite{mk}. 
Dashed line (red): the asymmetry calculated for the amplitude without the contributions of light neutral vector mesons 
$\phi$, the $c\bar c$ resonances ($J/\psi$, $\psi'$, \ldots), Bremsstrahlung, and the weak annihilation.}
\end{figure}

\begin{figure}[tb]
\begin{center}
\begin{tabular}{cc}
\includegraphics[width=5.8cm]{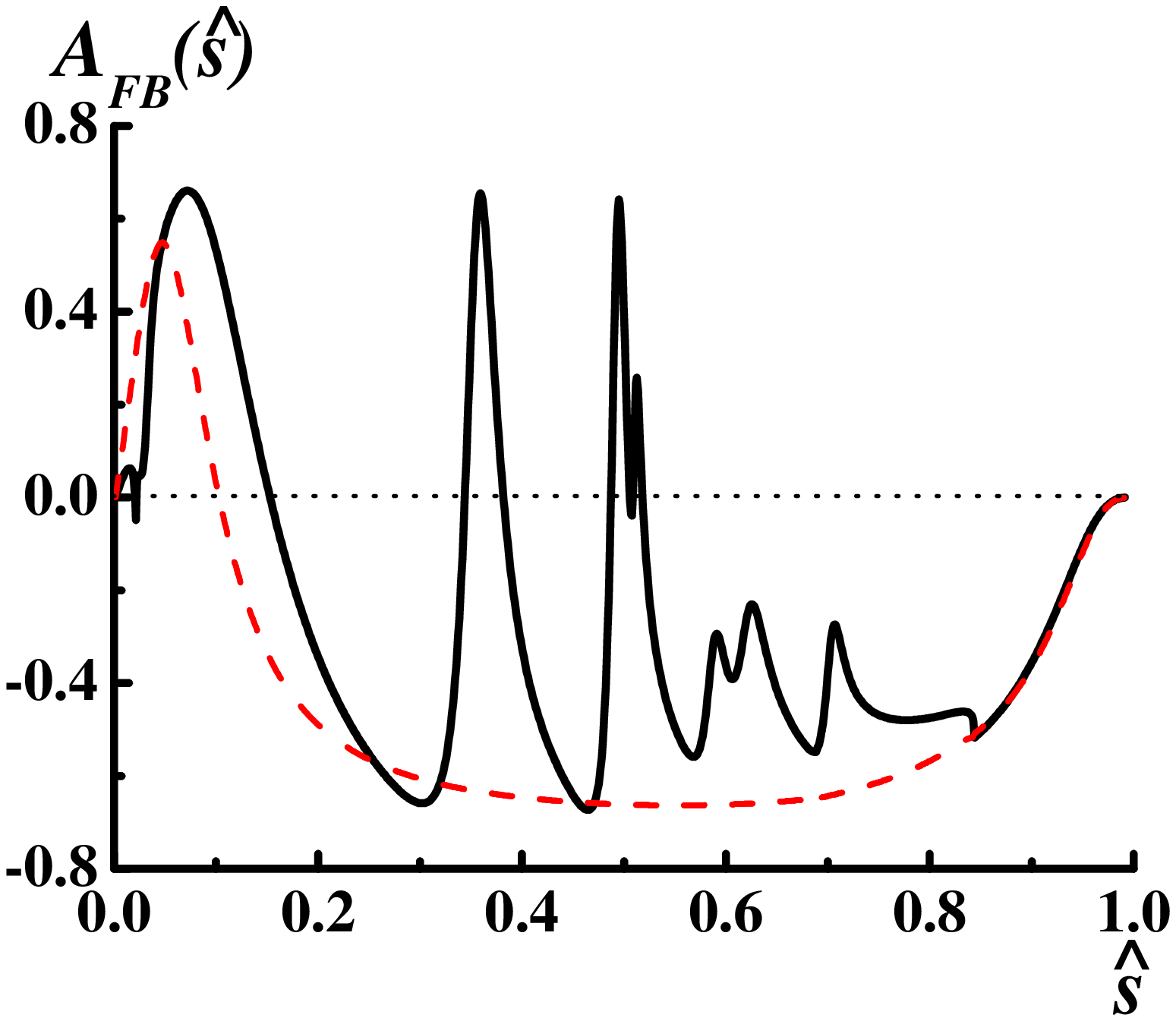} 
& 
\includegraphics[width=5.8cm]{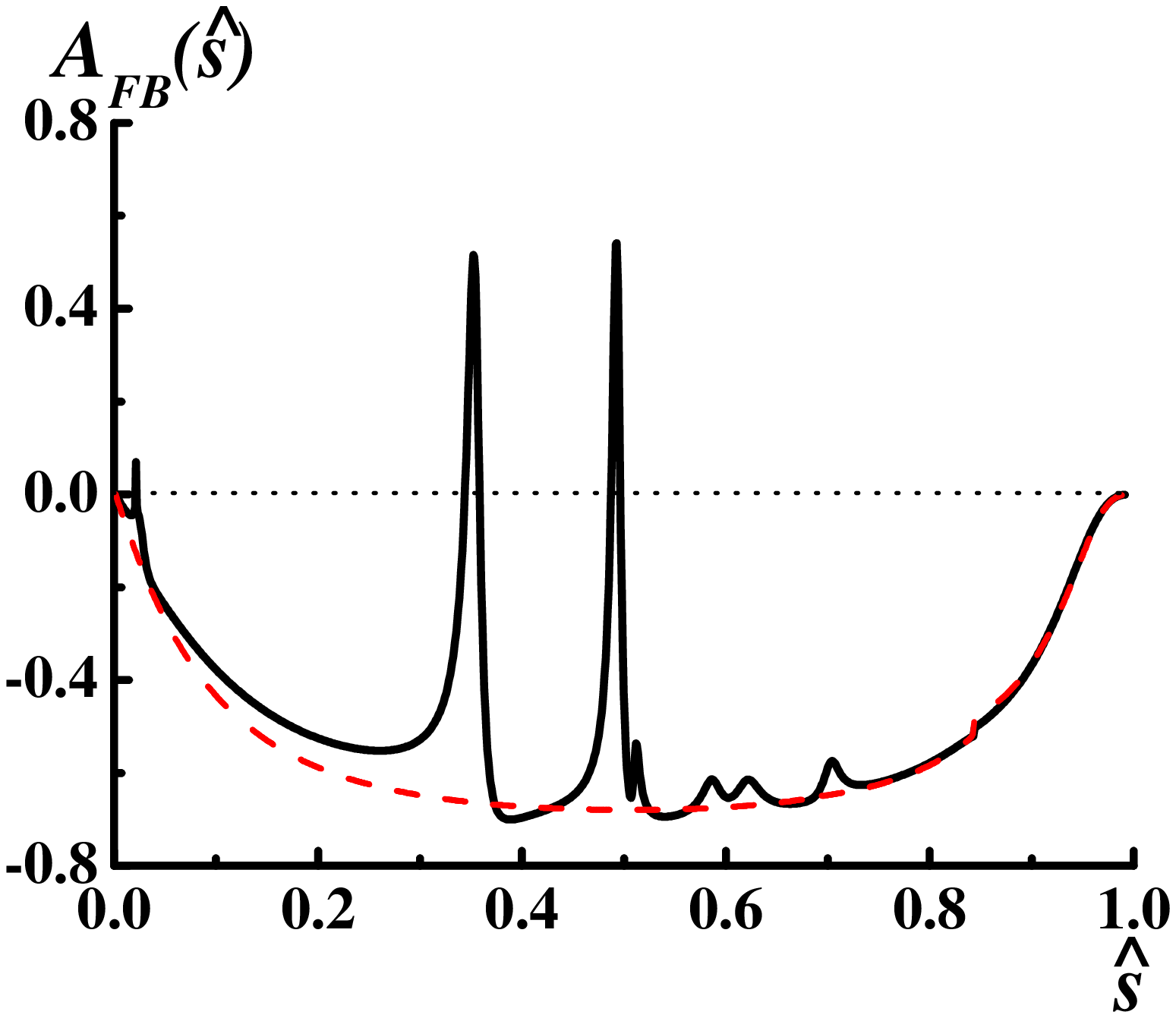} 
\\
(a) & (b)
\\
\includegraphics[width=5.8cm]{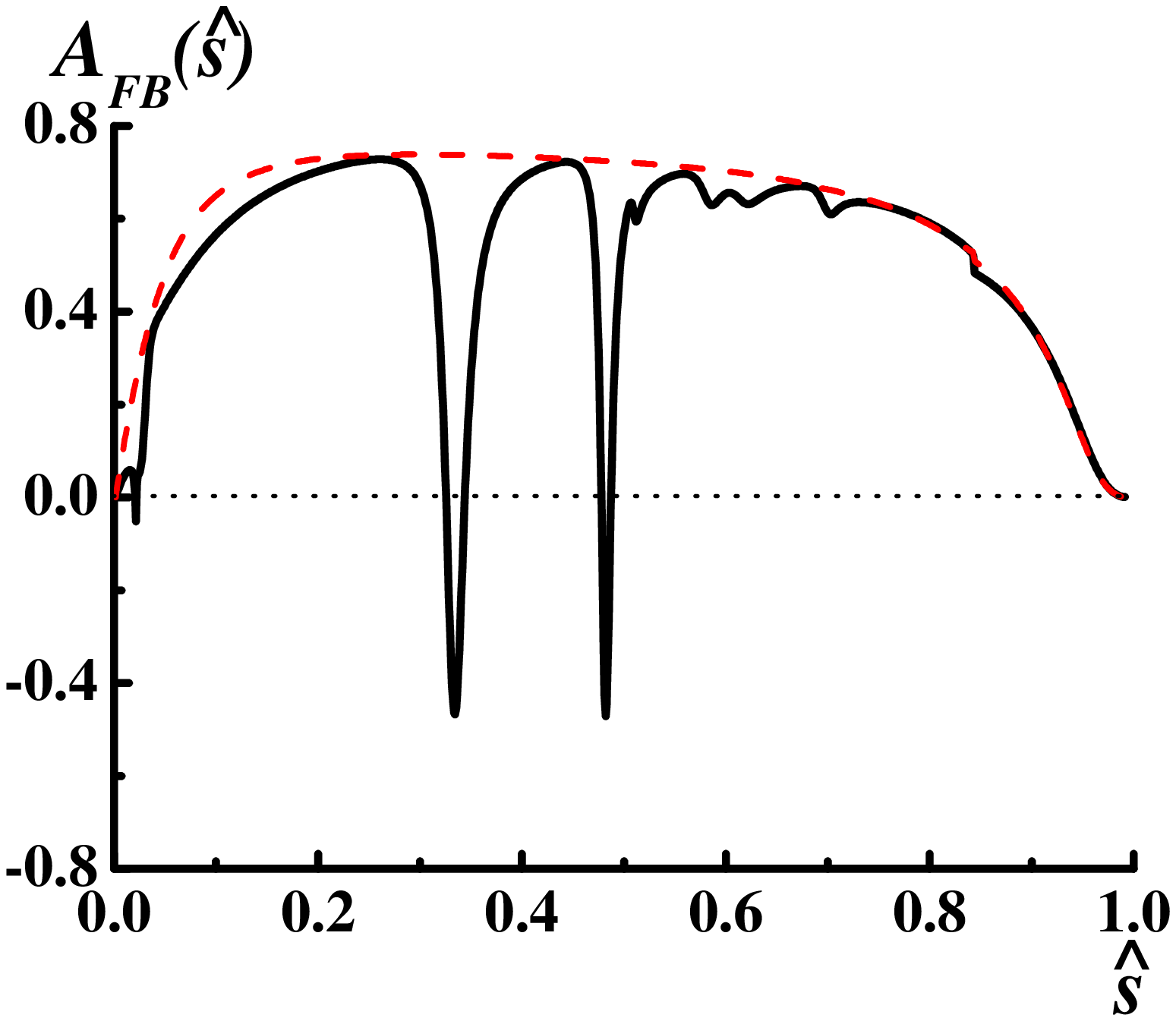} 
& 
\includegraphics[width=5.8cm]{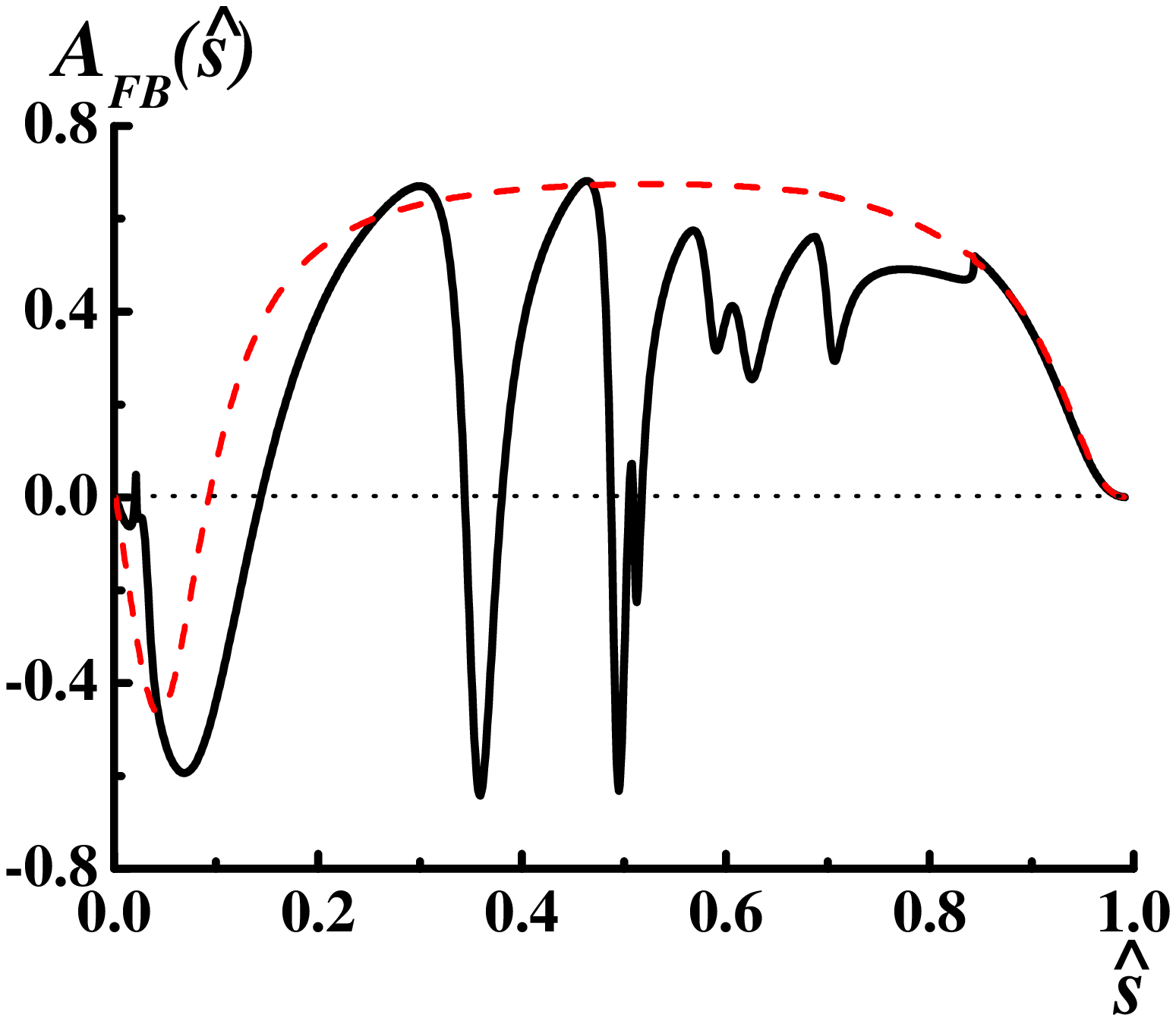}
\\
(c) & (d)
\end{tabular}
\end{center}
\caption{ \label{Fig:3}
$A_{FB}(\bar s)$ for $\bar B_d\to \gamma\mu^+\mu^-$ decays: 
(a) In the SM.  
(b) For $C_{7\gamma}=-C^{\rm SM}_{7\gamma}$. 
(c) For $C_{9V}=-C^{\rm SM}_{9V}$. 
(d) For $C_{10A}=-C^{\rm SM}_{10A}$. 
Solid line (black): the asymmetry calculated for the full amplitude of Ref.~\cite{mk}. 
Dashed line (red): the asymmetry calculated for the amplitude without the contributions of light neutral vector mesons 
$\omega$, $\rho^0$, the $c\bar c$ resonances ($J/\psi$, $\psi'$, \ldots) Bremsstrahlung, and the weak annihilation.}
\end{figure}

The calculated forward-backward asymmetries are presented in Figs.~\ref{Fig:1}--\ref{Fig:3}. 

The decay $\bar B_s\to \phi\mu^+\mu^-$ (Fig.~\ref{Fig:1}) is of special interest: 
the detector LHCb may accumulate sufficient 
data sample for this decay already after the first few months of operation. 
Qualitatively, the asymmetry has the same structure as in the 
$B\to K^*\mu^+\mu^-$ decays: its behaviour at small $\hat s$ 
is sensitive to the invertion of the signs of $C_{7\gamma}$ and $C_{10A}$ compared 
to the SM. Fig.~\ref{Fig:1} shows the influence of the sign invertion in one of the 
Wislon coefficients compared to the SM. 
      
Figs.~\ref{Fig:2} and \ref{Fig:3} present $A_{FB}$ for the radiative decays 
$\bar B_s\to \gamma\mu^+\mu^-$ and $\bar B_d\to \gamma\mu^+\mu^-$, respectively. 
Qualitatively, the asymmetry behaves at large and intermediate $\hat s$ similarly 
to the $B_s\to \phi\mu^+\mu^-$ decay, although has a larger magnitude. At small 
$\hat s$, however, the assymetry is influenced by the neutral light vector resonances 
$\phi$, $\omega$, and $\rho^0$ \cite{mk}: these resonances cause a strong distortion 
of the full asymmetry compared to a nonresonance asymmetry. In particular, they leads  
to a visible shift of the ``zero-point'' compared to its location in the non-resonant 
asymmetry, which may be reliably calculated in the SM \cite{null-test}. 

    
\subsection{CP-violating asymmetries}
We present the time-independent and the time-dependent CP-asymmetries in 
$B_d\to (\rho,\gamma) \mu^+\mu^-$. Concerning the $B_s\to (\phi,\gamma) \mu^+\mu^-$ decays 
we would like to mention the following: we have cal\-cu\-la\-ted these asymmetries and found 
that $A_{CP}(\hat s)$, mainly due to flavor oscillations of the $B_s$ mesons, 
is extremely small (smaller than 0.1\%) and therefore cannot be studied experimentally;  
$A_{CP}(\tau)$ is not small but measuring this asummetry would require time resolution much smaller 
than the $B_s$ lifetime. 

%
\subsubsection{Time-independent asymmetry}
\begin{figure}[!ht]
\begin{center}
\begin{tabular}{cc}
\includegraphics[width=7.2cm]{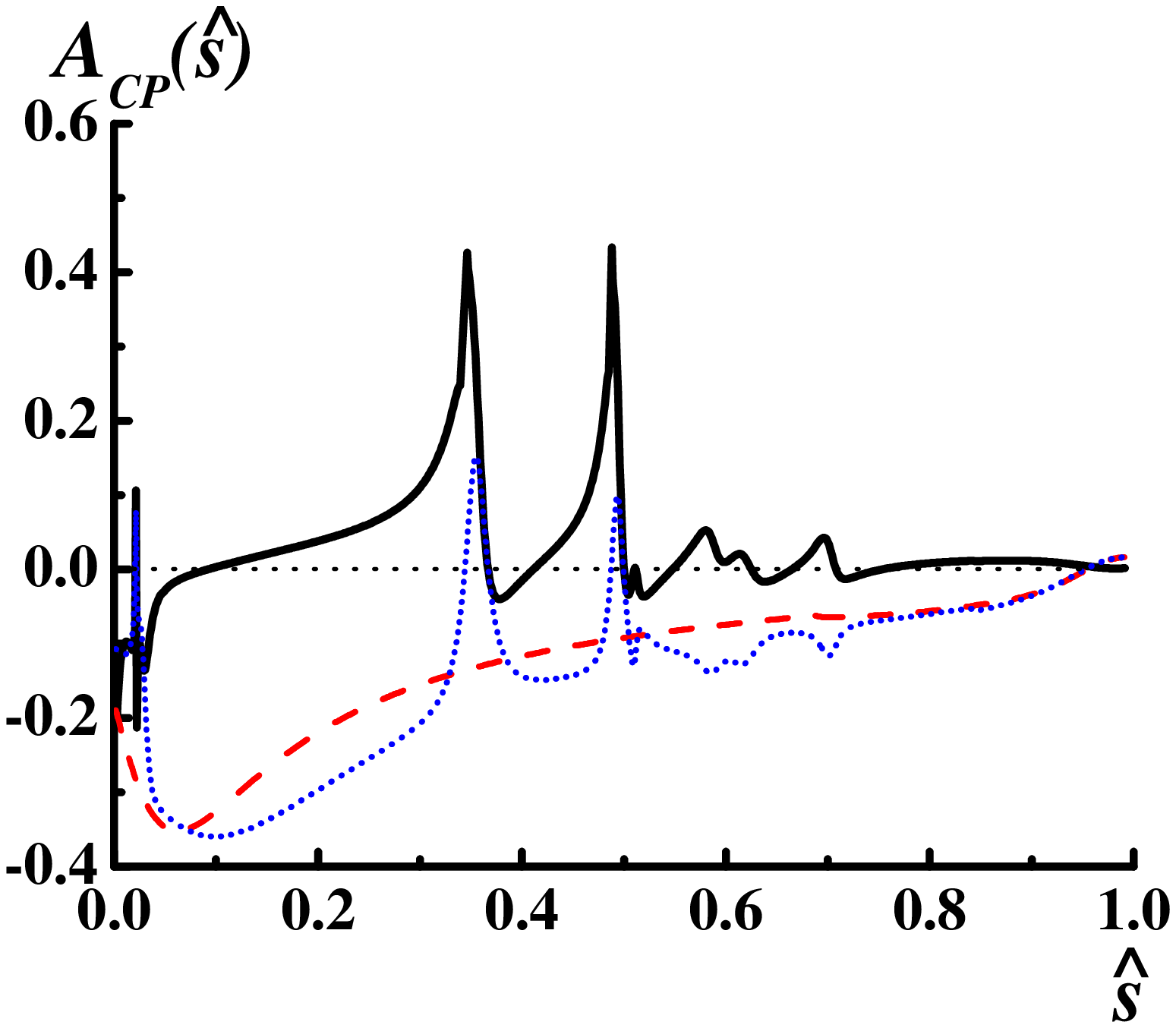} & 
\includegraphics[width=7.2cm]{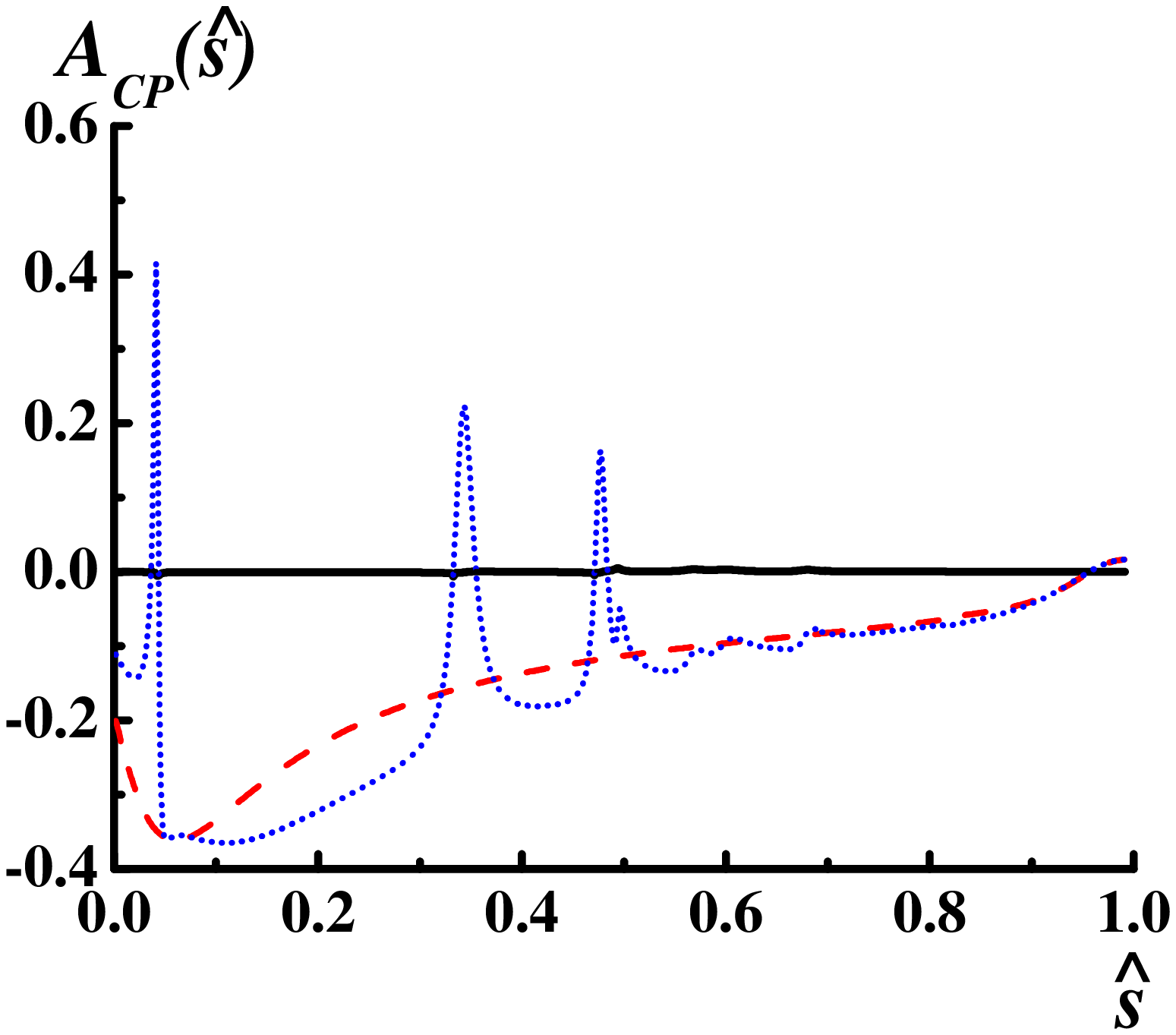} 
\\
(a) & (b)
\end{tabular}
\end{center}
\caption{\label{Fig:4}
The influence of $B$-meson flavor oscillations upon CP-violating asymmetries: 
(a) $B_d\to \gamma\mu^+\mu^-$, 
(b) $B_s\to \gamma\mu^+\mu^-$.
Dashed line (red): $A_{CP}$ without resonances and without flavor oscillations; 
Dotted line (blue): $A_{CP}$ with resonances but without flavor oscillations; 
Solid line (black): $A_{CP}$ after flavor oscillations have been taken into account.}
\end{figure}

First, we would like to demonstrate the impact of flavor oscillations of the initial mesons on the 
resulting CP-violating asymmetries.
Fig.~\ref{Fig:4} shows $A_{CP}(\hat s)$ for $B_{d,s}\to\gamma\mu^+\mu^-$ decays. Obviously, flavour oscillations 
lead to a strong suppression of the the resulting CP-violating asymmetries in $B_d$ decays and to a complete
vanishing of $A_{CP}$ in $B_s$ decays.  

\begin{figure}[!ht]
\begin{center}
\begin{tabular}{ccc}
\includegraphics[width=5.0cm]{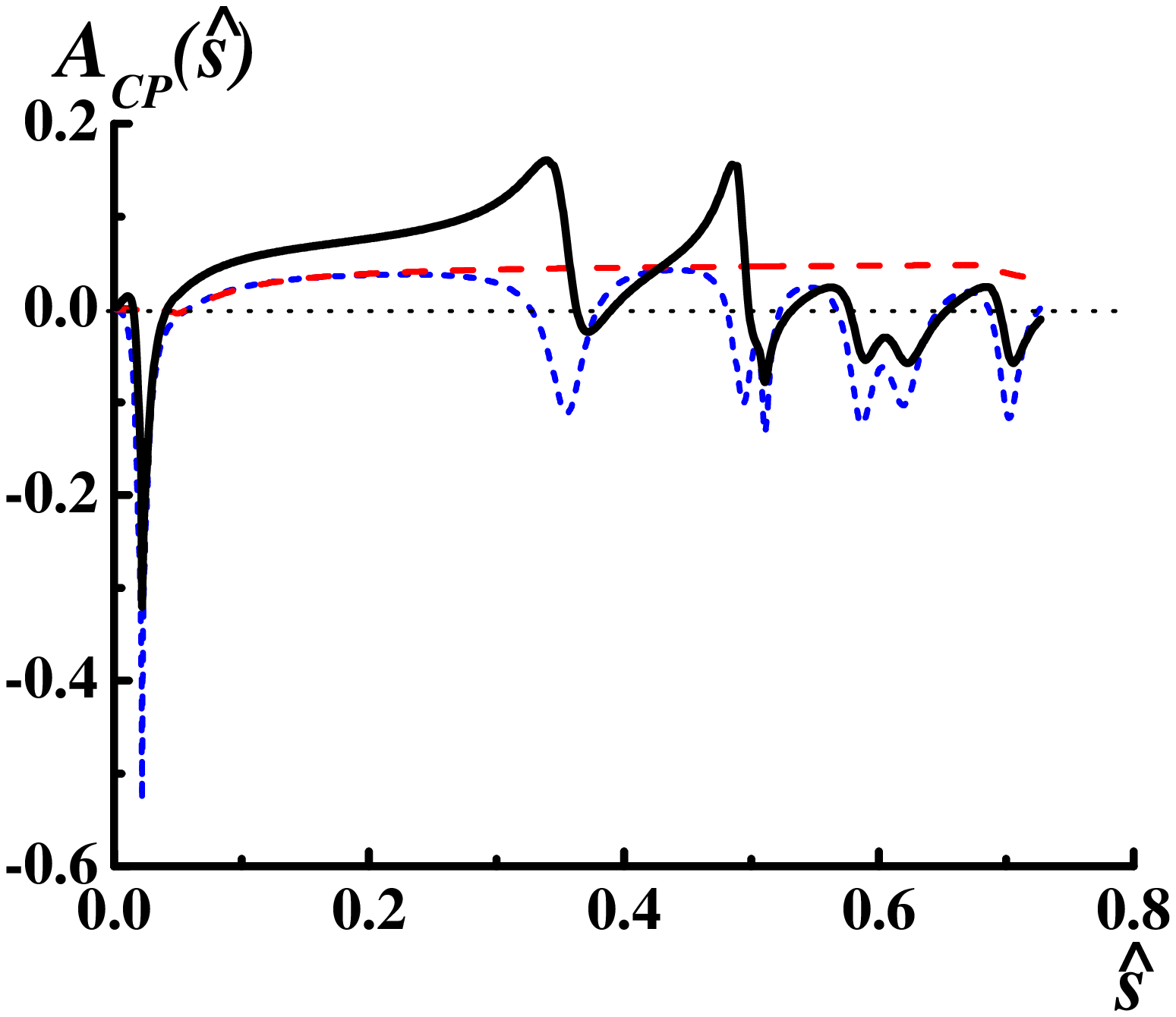} & 
\includegraphics[width=5.0cm]{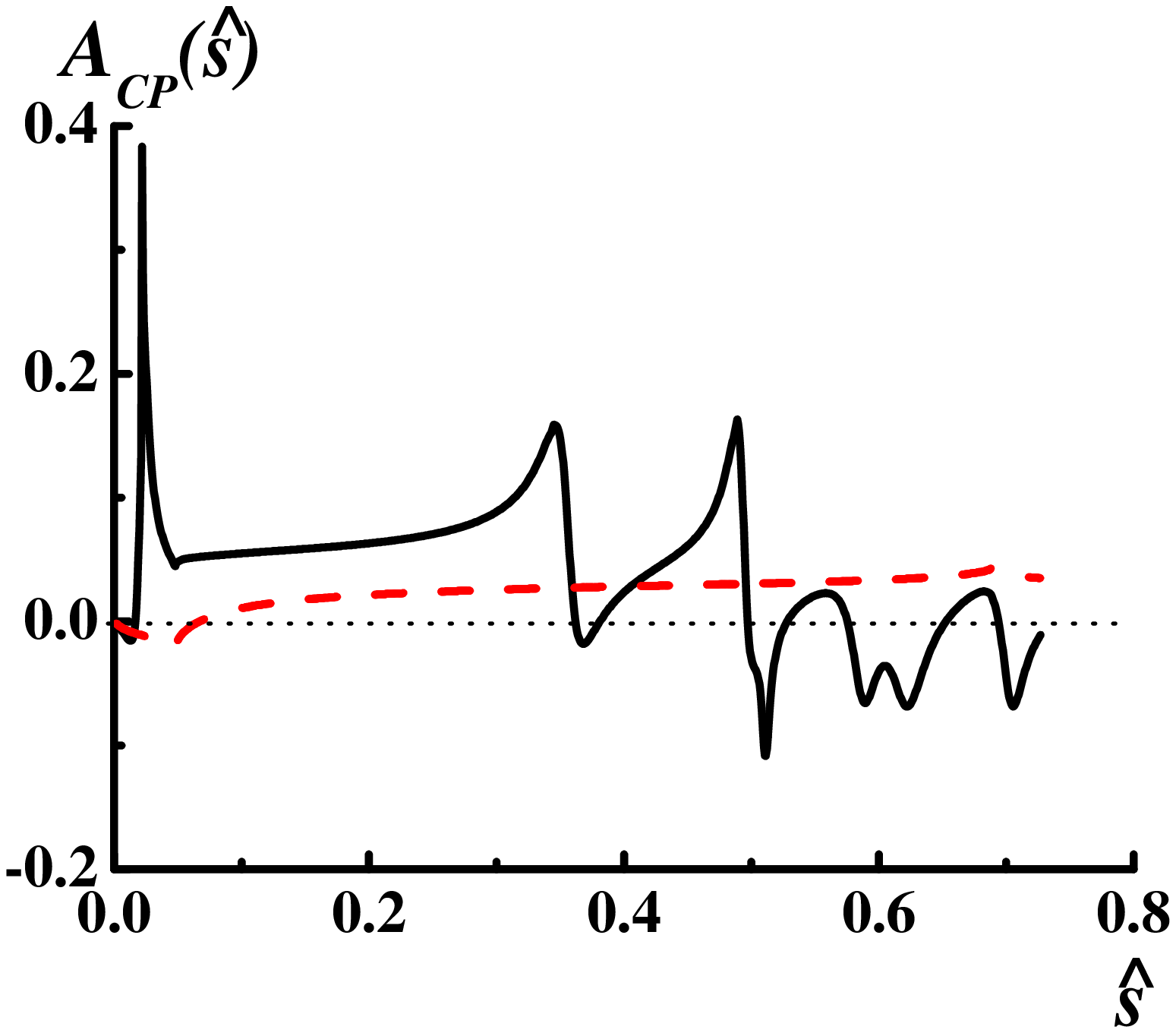} &
\includegraphics[width=5.0cm]{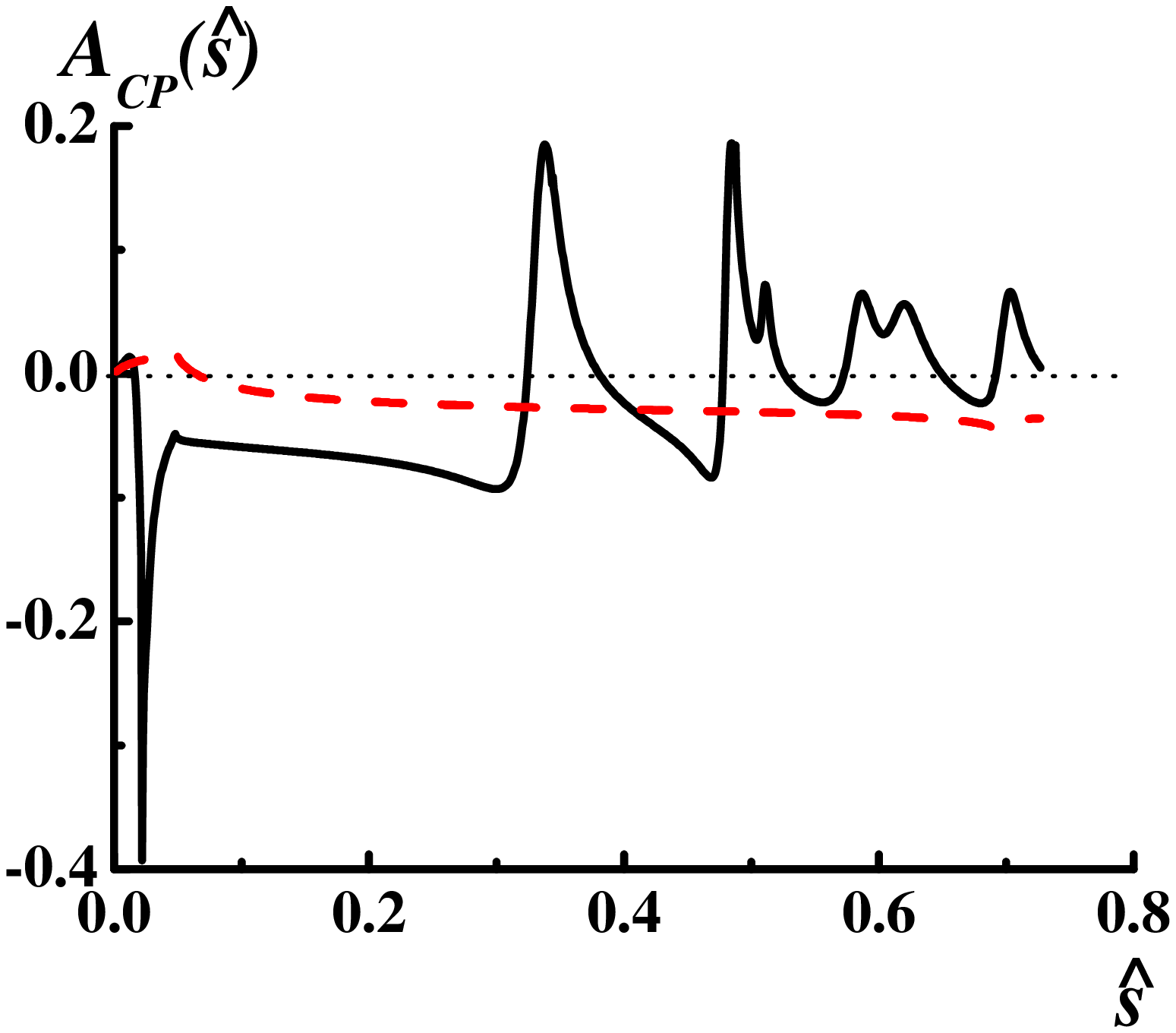} 
\\
(a) & (b) & (c)
\end{tabular}
\end{center}
\caption{\label{Fig:4.1}
Time-independent CP-asymmetry $A_{CP}(\hat s)$ in $B_d\to \rho\mu^+\mu^-$ decays.
(a) SM  (b) $C_{7\gamma}=-C^{\rm SM}_{7\gamma}$ (c) $C_{9V}=-C^{\rm SM}_{9V}$. Flavor oscillations have been taken into account. 
Solid line (black) line: full asymmetry. 
Dashed (red) line: nonresonant asymmetry.
Dotted (blue) line shows the asymmetry if flavor oscillations are not taken into account.}
\end{figure}

\begin{figure}[!hb]
\begin{center}
\begin{tabular}{ccc}
\includegraphics[width=5.0cm]{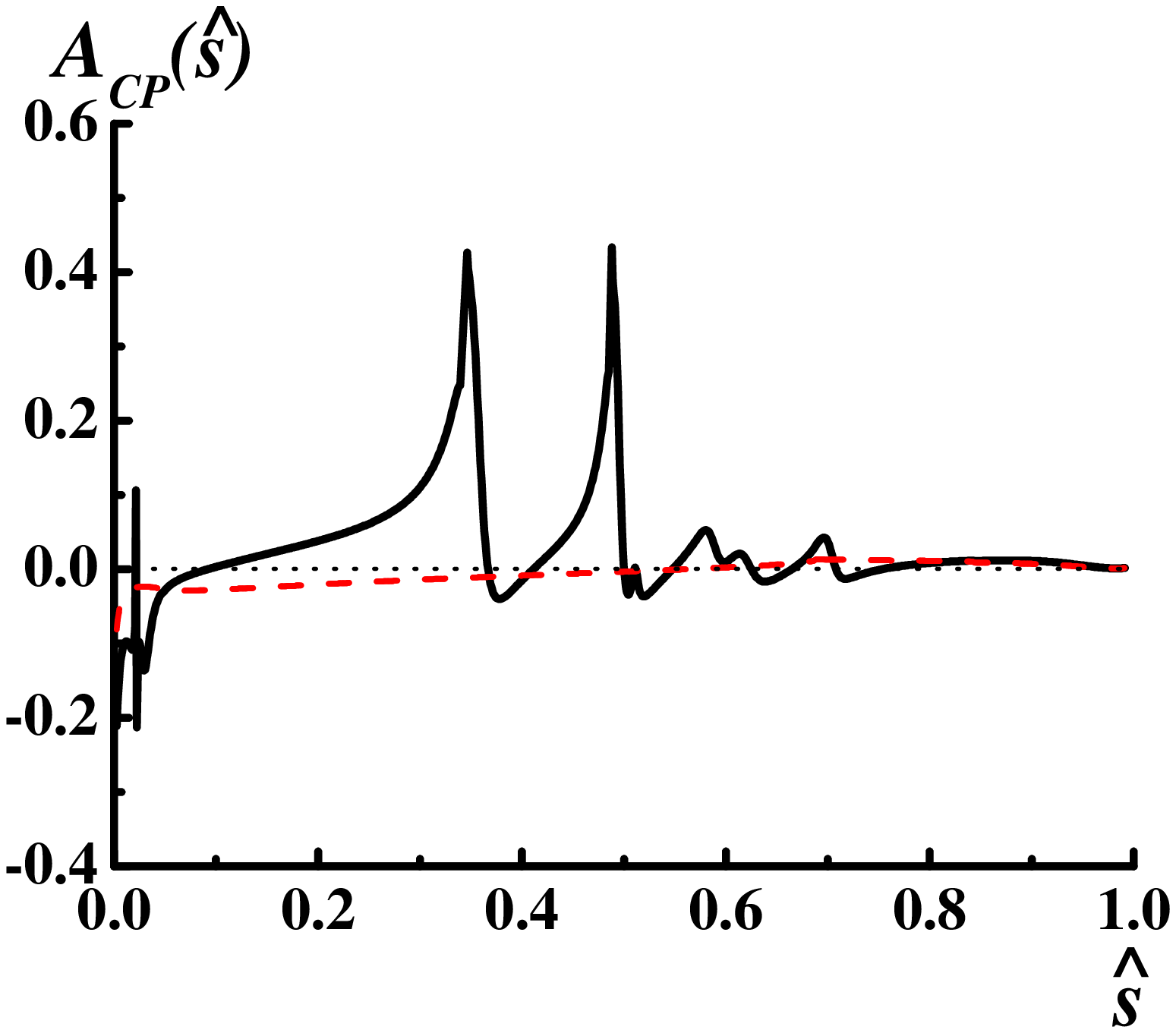} & 
\includegraphics[width=5.0cm]{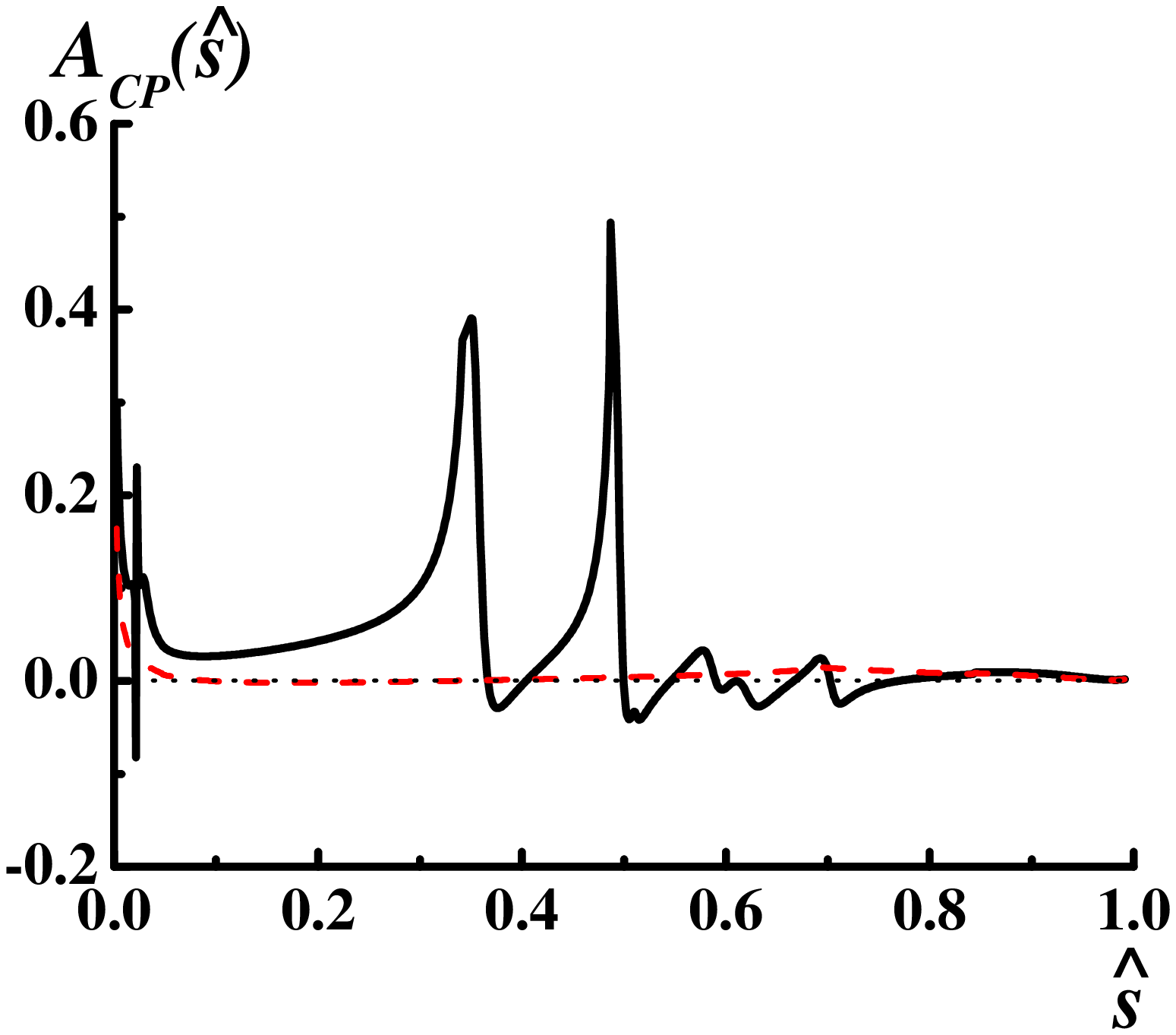} &
\includegraphics[width=5.0cm]{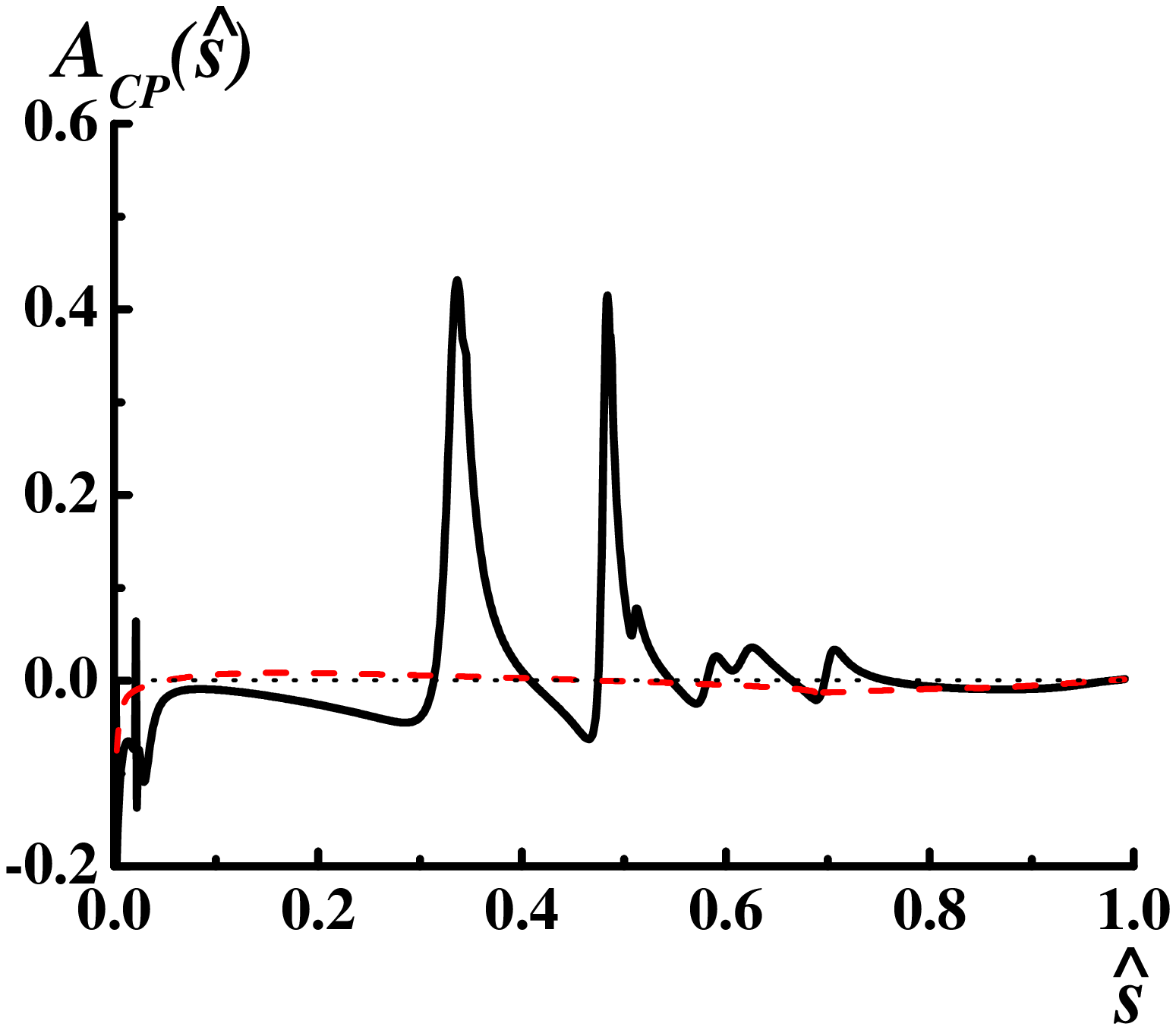}
\\
(a) & (b)& (c)
\\
\end{tabular}
\end{center}
\caption{\label{Fig:4.2}
Time-independent CP-asymmetry $A_{CP}(\hat s)$ in $B_d\to \gamma\mu^+\mu^-$ decays.
(a) SM  (b) $C_{7\gamma}=-C^{\rm SM}_{7\gamma}$ (c) $C_{9V}=-C^{\rm SM}_{9V}$. 
Solid (black) line: full asymmetry. Dashed (red) line: nonresonant asymmetry.
Flavor oscillations have been taken into account. 
}
\end{figure}

Figs. \ref{Fig:4.1} and \ref{Fig:4.2} display $A_{CP}(s)$ for $B_d\to \rho\mu^+\mu^-$ and 
$B_d\to \gamma\mu^+\mu^-$ decays, respectively. 

For $B_d\to \rho\mu^+\mu^-$ decays, the asymmetry reaches a 30-40\% level in the region of light vector resonances, and 
a level of 10\% between the light and $c\bar c$ resonances. Notice that flavor oscillations enhance the asymmetry by 
a factor 2.  

For $B_d\to \gamma\mu^+\mu^-$ decays the asymmetry is smaller and may be measured only in the region of light vector resonances. 

Both for $B_d\to \rho\mu^+\mu^-$ and $B_d\to \gamma\mu^+\mu^-$ decays the asymmetry is sensitive to the signs of the Wilson
coefficients $C_7$ and $C_9$. The asymmetry is however not sensitive to the invertion of the sign of $C_{10}$. 



\subsubsection{Time-dependent asymmetry}
\begin{figure}
\begin{center}
\begin{tabular}{ccc}
\includegraphics[width=5.0cm]{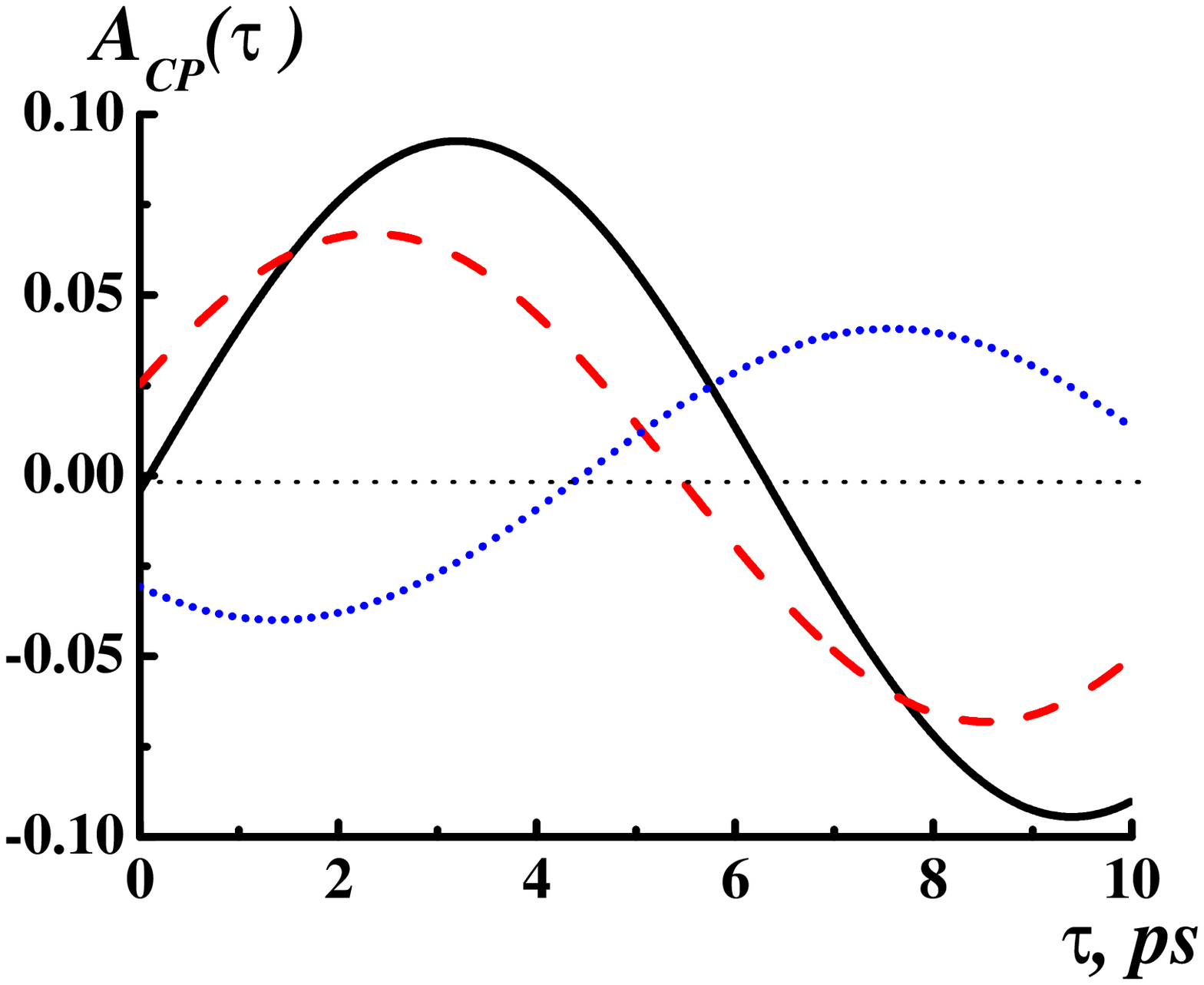} &
\includegraphics[width=5.0cm]{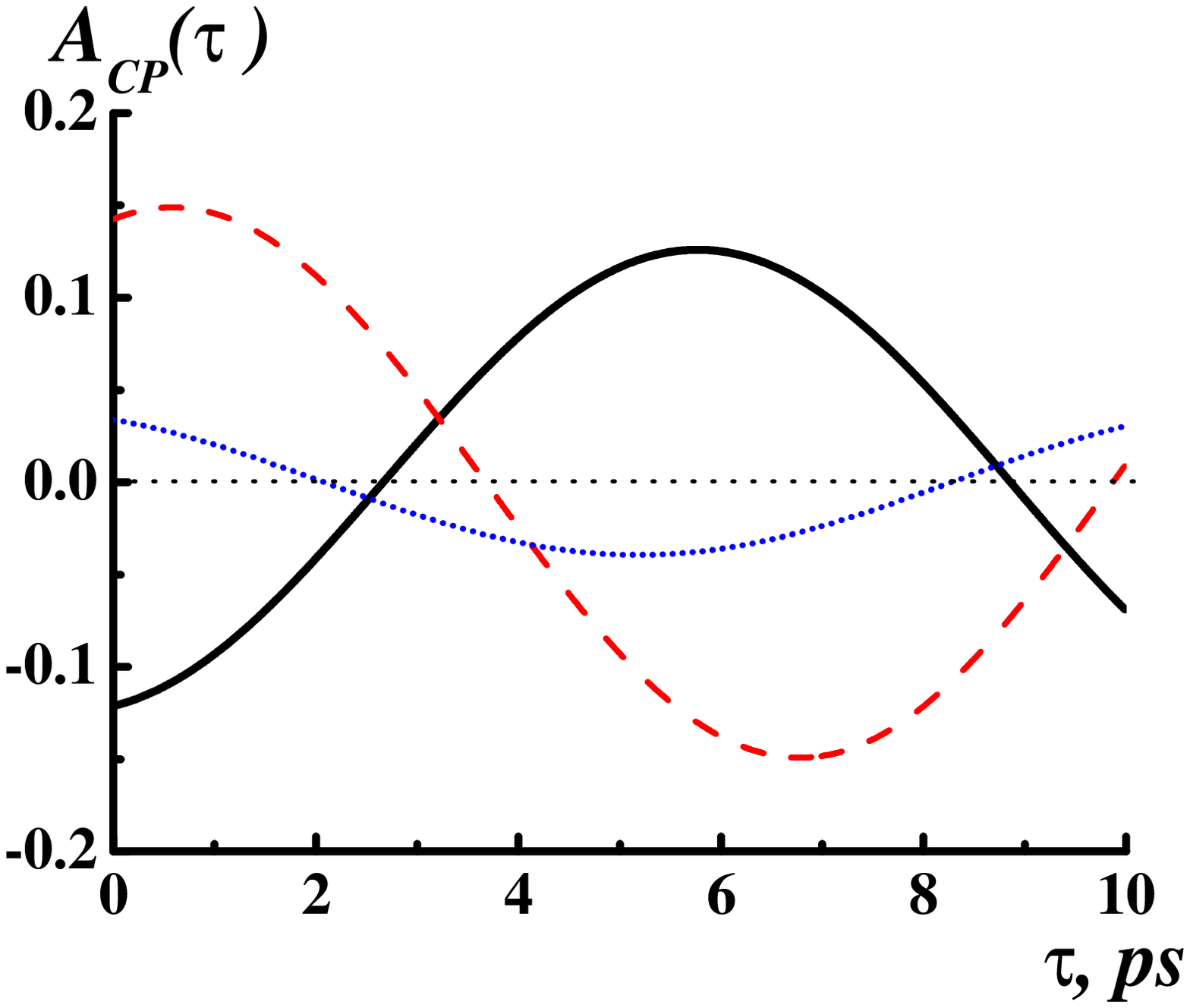} &
\includegraphics[width=5.0cm]{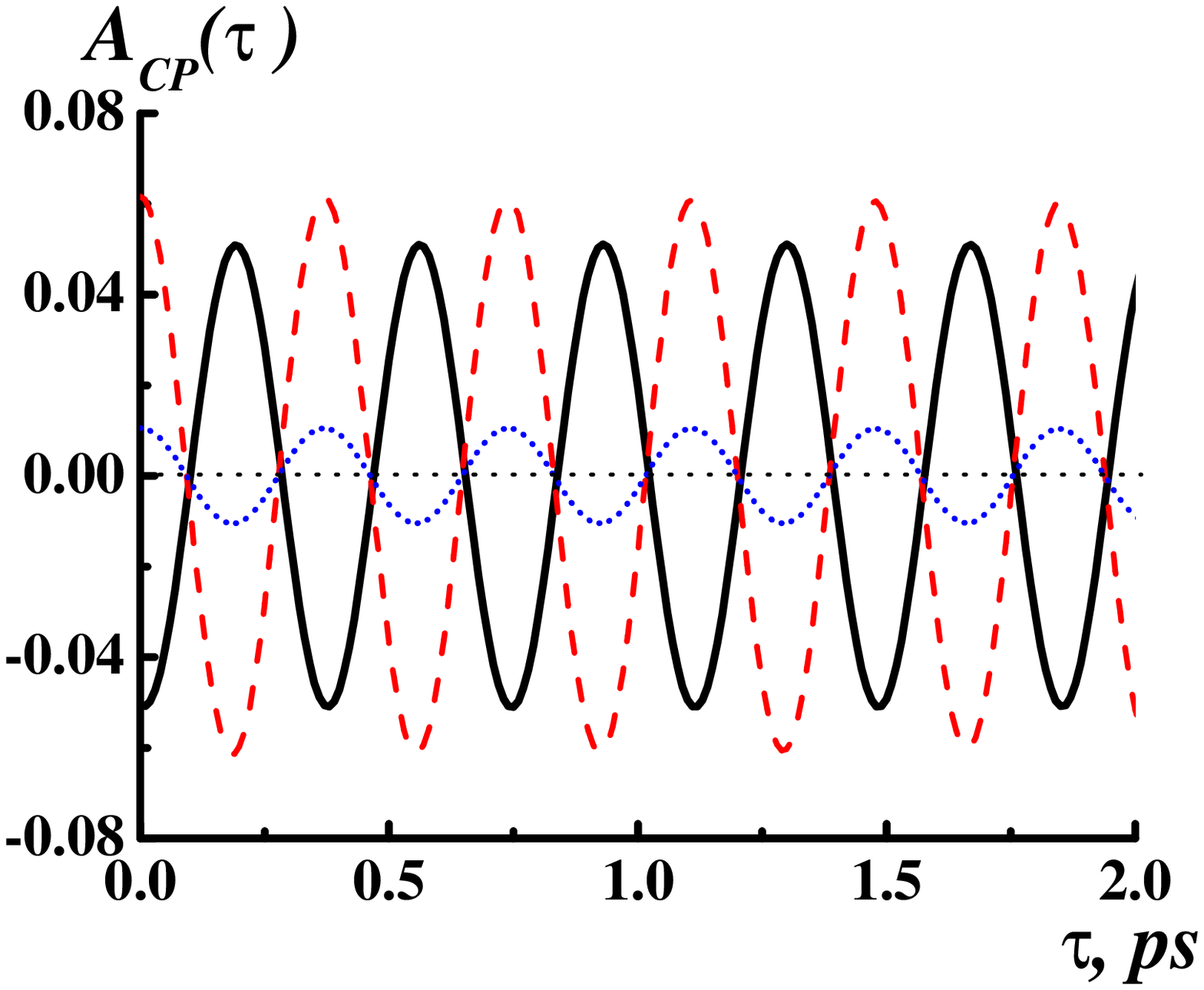} 
\\
(a) & (b) & (c)
\end{tabular}
\end{center}
\caption{\label{Fig:5}
Time-dependent asymmetry $A_{CP}(\tau)$: 
(a) $B_d\to \rho\mu^+\mu^-$; 
(b) $B_d\to \gamma\mu^+\mu^-$;  
(c) $B_s\to \gamma\mu^+\mu^-$. 
Solid line (black): SM. 
Dashed line (red): $C_{7\gamma}=-C^{\rm SM}_{7\gamma}$.  
Dotted line (blue): $C_{9V}=-C^{\rm SM}_{9V}$.}
\end{figure}
Fig.~\ref{Fig:5} plots the time-dependent asymmetry $A_{CP}(\tau)$ for 
$B_d\to \rho\mu^+\mu^-$ (a)  $B_d\to \gamma\mu^+\mu^-$ (b) and 
$B_s\to \gamma\mu^+\mu^-$ (c) decays. The region around the $J/\psi$ and $\psi'$ resonances $0.33\le \hat s \le 0.55$ 
was excluded from the integration 
while calculating the time-dependent asymmetries. This procedure corresponds to the analysis of the experimental data. 

The asymmetry in $B_d$ decays reaches a level of 10\% at the time-scale of a few $B$-meson lifetimes ($\tau_{B_d}$=1.53 ps) 
and may be studied experimentally. It also exhibits a sensitivity the the extentions of the SM. 


\section*{Conclusions}
We presented the analysis of the forward-backward and the CP-violating asymmetries in rare semileptonic 
and radiative leptonic $B$-decays. Our results may be summarized as follows: 

\begin{enumerate}
\item
We obtained the analytic results for the time-dependent and time-independent $CP$-asymmetries in rare radiative 
leptonic $B$-decays $B_{d,s}\to \gamma \ell^+\ell^-$. 

\item 
We presented numerical results for the forward-backward asymmetry in $B_{s}\to \phi \mu^+\mu^-$ decays which may 
be measured in the near future at the LHCb. This asymmetry, as could be expected, has a very similar shape to the 
asymmetry in $B_{d}\to K^* \mu^+\mu^-$ decays and thus may be used for ``measuring'' the signs of the Wilson coefficients 
$C_{7\gamma}$, $C_{9V}$, and $C_{10A}$. 

\item 
We studied the forward-backward asymmetry in $B_{d,s}\to \gamma \ell^+\ell^-$ decays taking into account the vector resonance 
contributions, the Bremsstrahlung, and the weak annihilation effects. 
We noticed that the light neutral vector resonances strongly distort the shape of the asymmetry at 
small values of the dilepton invariant mass. In particular, in the SM these resonances lead to a sizeable shift 
of the zero point of the full asymmetry compared to the zero-point of the non-resonant asymmetry. 
The $A_{FB}$ in this reaction reaches 60\% and thus may be studied experimentally at the LHC and 
the future Super-B factory. 

\item
We analysed the CP-violating asymmetries (both time-dependent and time-independent) in $B_d\to\rho\mu^+\mu^-$, 
$B_s\to\phi\mu^+\mu^-$, and $B_{s,d}\to\gamma\mu^+\mu^-$ decays. 

The asymmetries in $B_s$ decays are found to be very small 
and therefore to be of no practical interest. 

The asymmetries in $B_d$ decays reach measurable values and thus might provide additional tests of the SM and its extentions. 
These potentially interesting cases are: 
(i) The time-independent CP-violating asymmetry $A_{CP}(\hat s)$ in 
$B_{d}\to\rho\mu^+\mu^-$ decays 
in the region below $c\bar c$ resonances (10-30 \% level) 
and $A_{CP}(\hat s)$ in $B_{d}\to \gamma\mu^+\mu^-$ in the region of 
light neutral vector resonances (5-10 \% level). 
(ii) The time-dependent CP-violating asymmetry $A_{CP}(\tau)$ in $B_{d}\to(\rho,\gamma)\mu^+\mu^-$ 
decays (10\% level). 
\end{enumerate}


\section{Acknowledgments}
We are grateful to S.~Baranov, A.~Berezhnoj, V.~Galkin, Yu.~Koreshkova, W.~Lucha 
and Miu I.~D.~Mur for discussions and to G.~Hiller for comments on the initial version 
of the paper. The work was supported in part by grant for leading scientific schools 
4142.2010.2, by FASI state contract 02.740.11.0244, and by FWF project P20573. 



\end{document}